\documentclass[fleqn,twocolumn,tighten,apj]{aastex63} %

\usepackage{textcomp,amsmath,bm,nccmath,gensymb}

\usepackage{scalerel}

\usepackage{hyperref,multirow,dsfont,balance,enumitem,tabularx,booktabs}

\newcommand\HII{H\protect\scaleto{$II $}{1.2ex}}
\newcommand{\Lagr}{\mathcal{L}}
\DeclareMathSymbol{\shortminus}{\mathbin}{AMSa}{"39}

\shorttitle{Photometry on Structured Backgrounds}
\shortauthors{Saydjari and Finkbeiner}

\graphicspath{{./}{figures/}}

\begin{document}
\title{Photometry on Structured Backgrounds: Local Pixelwise Infilling by Regression}

\correspondingauthor{Andrew Saydjari}
\email{andrew.saydjari@cfa.harvard.edu}

\author[0000-0002-6561-9002]{Andrew K. Saydjari}
\affiliation{Department of Physics, Harvard University, 17 Oxford St., Cambridge, MA 02138, USA}
\affiliation{Harvard-Smithsonian Center for Astrophysics, 60 Garden St., Cambridge, MA 02138, USA}

\author[0000-0003-2808-275X]{Douglas P. Finkbeiner}
\affiliation{Department of Physics, Harvard University, 17 Oxford St., Cambridge, MA 02138, USA}
\affiliation{Harvard-Smithsonian Center for Astrophysics, 60 Garden St., Cambridge, MA 02138, USA}
\vspace*{-7mm}
\begin{abstract}

Photometric pipelines struggle to estimate both the flux and flux uncertainty for stars in the presence of structured backgrounds such as filaments or clouds. However, it is exactly stars in these complex regions that are critical to understanding star formation and the structure of the interstellar medium. We develop a method, similar to Gaussian process regression, which we term local pixelwise infilling (LPI). Using a local covariance estimate, we predict the background behind each star and the uncertainty on that prediction in order to improve estimates of flux and flux uncertainty. We show the validity of our model on synthetic data and real dust fields. We further demonstrate that the method is stable even in the crowded field limit. While we focus on optical-IR photometry, this method is not restricted to those wavelengths. We apply this technique to the 34 billion detections in the second data release of the Dark Energy Camera Plane Survey (DECaPS2). In addition to removing many $>3\sigma$ outliers and improving uncertainty estimates by a factor of $\sim 2-3$ on nebulous fields, we also show that our method is well-behaved on uncrowded fields. The entirely post-processing nature of our implementation of LPI photometry allows it to easily improve the flux and flux uncertainty estimates of past as well as future surveys.

\end{abstract}

\section{Introduction} \label{sec:Intro}

Upcoming projects such as the Legacy Survey of Space and Time (LSST) at the Vera C. Rubin Observatory \citep{jones2020scientific} and observations from the Nancy Grace Roman Space Telescope \citep{Akeson:2019:arXiv:} will deliver measurements of stellar flux (brightness) for billions of stars, including those near the Galactic plane. Measurement of stellar flux in the Galactic plane is complicated by crowding (which makes it difficult to partition flux between nearby overlapping sources), and the presence of structured backgrounds\footnote{We refer to filaments of gas and dust as ``background'' regardless of whether those features are physically in front of or behind the star.} (the flux of which should not be attributed to point sources).

The difficulties of crowding and structured backgrounds near the Galactic plane were explicitly noted in the Sloan Extension for Galactic Underpinnings and Evolution (SEGUE) \citep{Newberg:2003:AAS:}, the Sloan Digital Sky Survey (SDSS) release at low galactic latitudes \citep{finkbeiner2004sloan}, and the Dark Energy Camera Plane Survey (DECaPS) \citep{Schlafly:2018:ApJS:}. Some of these projects created modified photometric pipelines to better handle crowded fields and source deblending, but did not address structured backgrounds. Other works addressing photometry on structured backgrounds focus solely on methods for source identification \citep{Molinari:2011:A&A:} or use polynomial fits of varying degrees to model the true background, both of which complicate estimating the uncertainty on the reported stellar flux \citep{Traficante:2015:A&A:}. A general technique able to handle both biases and uncertainties resulting from structured backgrounds could not only improve future measurements in the galactic plane (Rubin and Roman), but also create value-added catalogues using images from any of the large-scale photometric surveys that target the Galactic plane such as SDSS, DECaPS, WISE \citep{Schlafly:2019:ApJS:}, the Two Micron All Sky Survey (2MASS) \citep{Skrutskie:2006:AJ:}, Pan-STARRS1 \citep{magnier2020pan}, GLIMPSE \citep{Benjamin:2003:PASP:}, VISTA Variables in the Via Lactea (VVV) \citep{minniti2010vista}, and the UKIRT Infrared Deep Sky Survey (UKIDSS) \citep{Lawrence:2007:MNRAS:}.

\begin{figure*}[!ht]
\centering
\includegraphics[width=\linewidth]{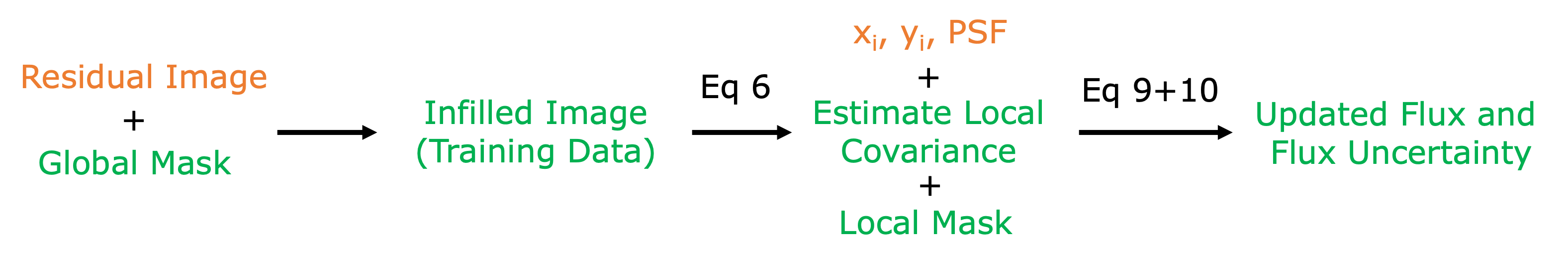}
\caption{A scheme illustrating the steps used in this work to improve flux and flux uncertainty estimates for sources on structured backgrounds. Orange text represents data products from a previous photometric solution for the image of interest. Green text represents data products generated as part of this work.}
\label{fig:scheme}
\end{figure*}

Instead of using classic background models that are smooth on a spatial averaging scale that is (and must be) larger than the Point Spread Function (PSF), we can instead perform an interpolation of the background behind the star. Using that local background model, we can separate the true flux of the star from background flux. In the machine-learning literature, reconstructing an unknown portion of an image is often called a ``conditional infilling'' or ``inpainting'' problem and many approaches exist, often using some version of generative adversarial networks (GANs) \citep{pathak2016context,ulyanov2018deep,wu2018conditional,van2018compressed}. In addition to deriving corrections to the flux as a result of the true background, it is important to quantify the overall uncertainty in the reported flux. Without reliable uncertainty estimates, any downstream statistical inference, which is the main application of these astronomical catalogues, is polluted with overly tight constraints. This is a limitation for approaches based on neural networks because it is difficult to obtain reliable uncertainty estimates on their predictions.

We improve the estimates of stellar flux and error bars on structured fields by solving the infilling problem behind the star \textbf{and} obtaining an uncertainty on that infill. To do this, we perform a regression for the unknown pixels given the values of neighboring pixels that are known. The local pixelwise infilling (LPI) used here is similar to Gaussian Process Regression (GPR), often called Gaussian Process Interpolation or Kriging. In GPR, the value of a function (the true background) is predicted at unobserved points conditioned on nearby observed points using a covariance function (called a kernel). Note that GPR assumes weak stationarity--that the kernel depends only on the (possibly angle-dependent) separation between points. Exact stationarity is often violated, but holds approximately in real applications. GPR is a form of nonparametric regression since each unobserved point is predicted without a prior specification on the functional form (linear, quadratic, etc.) of the interpolated image. However, GPR almost always uses a parametric kernel function. The benefit of parametric kernels is that the covariance can be sampled at any arbitrary separation. The downside is that an optimization over kernel parameters is required and one must worry about whether the parametric kernel is a good model of the true covariance of the data.

In astronomy, GPR has already found numerous applications, including foreground removal \citep{mertens2018statistical,Ghosh:2020:MNRAS:,Kern:2021:MNRAS:,Soares:2021:MNRAS:}, foreground modeling by point source interpolation \citep{Pinter:2018:IAUS:}, interpolating galactic-scale fields \citep{Platen:2011:MNRAS:,Yu:2017:PhRvD:,Leike:2019:A&A:,Williams:2022:MNRAS:}, cleaning/interpolating missing data \citep{czekala2015constructing, Baghi:2016:PhRvD:,Zhang:2021:AJ:,Ndiritu:2021:MNRAS:}, predicting photometric redshifts \citep{Way:2009:ApJ:,Almosallam:2016:MNRAS:}, lightcurve analysis \citep{evans2015uniform,littlefair2017robust,mcallister2017using,Angus:2018:MNRAS:}, radial velocity analysis \citep{rajpaul2015ghost,barclay2015radial,haywood2014planets}, removing instrumental systematics \citep{gibson2012gaussian,junklewitz2016resolve}, interpolating atmospheric turbulence to improve astrometry \citep{Fortino:2021:AJ:,Leget:2021:A&A:}, modeling PSF variations \citep{Gentile:2013:A&A:}, or predicting object properties \citep{bu2020estimation,Fielder:2021:MNRAS:}. 

The work of \cite{Fortino:2021:AJ:} notably improves astrometry (measurements of the position of stars) on images from the Dark Energy Camera (DECam), which is the same camera used for validating our photometric model in Section \ref{sec:Survey}. The extraction of clean foreground models by interpolating to remove point sources or bad pixels shown in \cite{Pinter:2018:IAUS:} and \cite{Zhang:2021:AJ:} serve as a proof of concept for this work. We further derive the correction to and uncertainties on the fluxes of those point sources as a result of the background model. Because we need to sample the covariance only on a grid (fixed by the pixelization of the camera) and need only a small number of points for the local background, we are able to use more flexible nonparametric models for the LPI covariance in contrast to the astronomical applications of GPR above.


\section{Background Model} \label{sec:Model}

Ideally, we would like to be able to estimate the background underneath a star from a detailed analysis of the nearby background. We find that we can successfully estimate the background and its uncertainty using an infilling technique that will be the focus of Section \ref{sec:LPI}. This technique dramatically improves flux and flux uncertainty estimates for stars in images of fields with significant nebulosity (Section \ref{sec:Neb}). In the intervening sections, we provide detailed descriptions, simulations, and validations of this technique, but the eager reader may wish to skip to Section \ref{sec:Survey}. Figure \ref{fig:scheme} broadly outlines the steps used in our technique.

\subsection{Likelihood} \label{sec:Like}

Photometric pipelines process astronomical images in order to model the location and flux (brightness) of the stars in the image. Focusing on surveys at low galactic latitudes where stars far outnumber galaxies, we set aside modeling galaxies for now. In addition to stars which are effectively point sources, these images often contain diffuse components such as emission from interstellar gas or reflection off of interstellar dust \citep{Magakian:2003:A&A:}. For example, the bright H$\alpha$ Balmer line of hydrogen and various emission lines for oxygen in different ionization states fall in the optical wavelength range \citep{lide2004crc}. Dense interstellar dust can also block background starlight, causing a decrease in detector counts over an extended region \citep{DiFrancesco:2002:AJ:}. These diffuse components and background sky counts are often combined into a single background term $b$, which when added to the sum of $N_{\rm{star}}$ point sources yields the true image $I$. The contribution of a point source to the image is modeled as the flux of the star (a scalar) times a Point Spread Function (PSF), which is the spatial distribution of photons arriving at the focal plane from a point source. The PSF deviates from a single point because of atmospheric turbulence, telescope optics, and sensor imperfections. Thus the true image ($I$)

\begin{align} \label{eq:imModel}
I(x,y) = b(x,y) + \!\! \sum_{i=1}^{N_{\rm{star}}}{f_i \times p(x\shortminus x_i,y\shortminus y_i)} + \epsilon(x,y)
\end{align}

where $b$ is the background, $f_i$ is the flux of star $i$, $p(x-x_i,y-y_i)$ is the PSF centered at position $(x_i,y_i)$ in the image, and $\epsilon(x,y)$ is the measurement noise of each pixel. Then the residual between the model and the true image is $r = I-\hat{I}$, where we have dropped the explicit $(x,y)$ dependence from here on.\footnote{In the code released with this paper, we use the opposite convention for the residuals, $r = \hat{I}-I$.}

\begin{align} \label{eq:residual}
r =  \epsilon + b - \hat{b} + \sum_{i=1}^{N_{\rm{star}}}{f_i \ p(x_i,y_i)} - \sum_{i=1}^{\hat{N}_{\rm{star}}}{\hat{f_i} \ \hat{p}(\hat{x_i},\hat{y_i})}
\end{align}

The noise in observed detector counts is from a mix of Poisson (photon events, shot noise) and Gaussian (read noise) processes. However, in optical surveys, we are generally in the large count limit and thus approximate the observed noise in each pixel as Gaussian. Then, we find the model parameters that minimize the negative log-likelihood.\footnote{In what follows, we will suppress a factor of 2 and normalization constants when writing likelihoods since they do not depend on model parameters and thus do not affect the optimization.}

\begin{ceqn}
\begin{align} \label{eq:resLikeli}
-2 \ln \Lagr (\hat{f_i}, \hat{x_i}, \hat{y_i}, \hat{b}) = \ln(\det(2\pi \hat C)) + r^T \hat C^{-1} r \
\end{align}
\end{ceqn}

To solve this problem in full generality, one has to optimize the likelihood with respect to $\hat{N}_{\rm{star}}$, a process known as transdimensional inference \citep{Brewer:2013:AJ:,Portillo:2017:AJ:,Feder:2020:AJ:,Liu:2021:arXiv:}. Such an approach can be made Bayesian by including a prior on $\hat{N}_{\rm{star}}$ and the other parameters. In practice, a heuristic deblending code that finds peaks in the image, assigns sources to those locations, and estimates a parametric PSF model is often used. Then, the optimization is performed only over $\hat{f}_i$, $\hat{x}_i$, and $\hat{y}_i$ with the other model parameters held fixed \citep{Stetson:1987:PASP:,Bertin:1996:A&AS:,Schlafly:2018:ApJS:}. We consider the latter approach. In what follows, we will also ignore uncertainties and corrections resulting from estimation of the stellar position ($\hat{x}_i$ and $\hat{y}_i$) such as those outlined in \cite{Portillo:2020:AJ:} and any errors resulting from PSF mismodeling, focusing instead on the previously unaddressed uncertainties and corrections to $\hat{f}_i$ in the presence of mismodeled backgrounds. 

It is tempting to assume that $r$ is mean zero (as assumed in Equation \ref{eq:resLikeli}) with $\hat C$ diagonal (as is often done), which would be nearly true in the limit of perfect modeling and noisy data. Then, the diagonal of $\hat C$ is simply the variance of each pixel (proportional to model counts assuming Poisson noise), and matrix inversion can be computed with little computational expense as element-wise division. However, the $\hat{b}$ used in photometric pipelines is often not an accurate model of the background $b$ and in this work we will consider how to deal with a non-diagonal $C$ resulting from mismodeled backgrounds.

\subsection{Literature Background Models} \label{sec:Back}

Popular photometric pipelines such as \textsc{daophot} \citep{Stetson:1987:PASP:}, SExtractor \citep{Bertin:1996:A&AS:}, the Tractor \citep{Lang:2016:ascl:}, and \texttt{crowdsource} \citep{Schlafly:2018:ApJS:} use background models that are smooth over large (relative to the PSF) spatial scales. Prior works have explored different methods to obtain a robust and/or unbiased estimation of a constant background value for a subset of an image. Most of these methods operate on the histogram of pixel values in the subimage and estimate the mean, median, or mode, often with iterative outlier removal (``$\sigma$-clipping'') \citep{DaCosta:1992:ASPC:}. The final background model applied at the pixel scale is then a smooth interpolation of individual subimage estimations. These approaches necessarily have an averaging scale (often ~32-128 pixels) set by the size of the subimages considered \citep{Bertin:1996:A&AS:}. So $\hat{b}$ can be a good model of background components that vary on a spatial scale larger than the averaging scale. However, high resolution imaging, from the Hubble Space Telescope for example, indicates structures in the ISM at scales all the way down to $< 1$ arcsec, which would not be well-captured by $\hat{b}$. 

\begin{figure*}
\centering
\includegraphics[width=\linewidth]{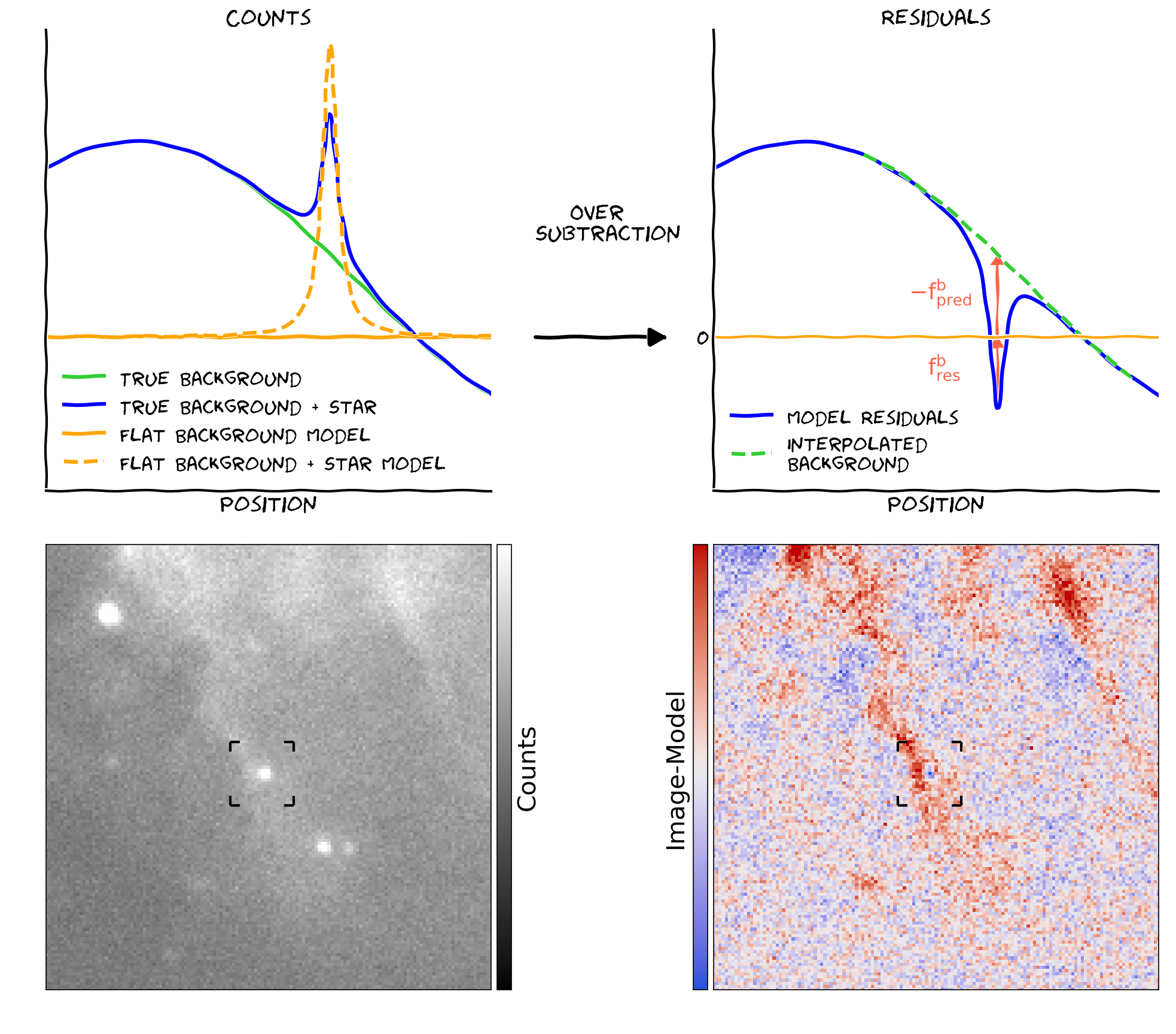}
\caption{\textit{Top Left}: Schematic showing overestimation of flux for a star (yellow dashed) given a flat background model (yellow solid) in the presence of a structured background (green solid). \textit{Top Right}: The residuals after subtraction which illustrate that $f^{\rm{b}}_{\rm{res}}$ accounts for the deviation of the residuals from zero and $-f^{\rm{b}}_{\rm{pred}}$ accounts for the deviation of the predicted background (green dashed) from zero.  \textit{Bottom Left}: An example star imposed on a structured background in a r-band image from DECaPS2 (Section \ref{sec:Survey}). \textit{Bottom Right}: Residuals after photometric fit of image at left using \texttt{crowdsource}. The fit drives the residual in a neighborhood around the highlighted source to be mean zero, relative to an incorrect background model, causing the center to be significantly oversubtracted.}
\label{fig:correctFactScheme}
\end{figure*}

In this case, the incorrect assumption that $r$ is mean zero can bias the flux estimates $\hat{f}_i$ for stars in the region. For example, if the observed background is above the model ($(b-\hat{b}) > 0$), then the $\hat{f}_i$ will overestimate the stellar flux so that $r$ is closer to mean zero (Figure \ref{fig:correctFactScheme}). The assumption that $r$ has a diagonal $C$ ignores nonzero (and as we will show in Section \ref{sec:SynVal} often large) off-diagonal contributions to the flux uncertainty, which are a result of the spatially correlated residuals from the mismodeled background.

\subsection{Local Pixelwise Infilling} \label{sec:LPI}

Since we cannot see behind the star, we describe the distribution of possible backgrounds by the set of images which are shifted horizontally/vertically by a number of pixels. We can view this as data augmentation of our single corrupted sample, where we have assumed local translation invariance of the statistics of the ISM (stationarity). This translation invariance should hold if the samples are ``local enough.'' We have \textbf{not} assumed rotational invariance (isotropy) because of the presence of highly directional filaments in the ISM (see Appendix \ref{sec:station}). Then, we perform a regression for the unknown pixels given the values of neighboring pixels that are known. Together, we term this approach local pixelwise infilling (LPI).

This regression can be viewed as GPR with a non-parametric anisotropic kernel, so long as the covariance matrix we estimate is actually stationary. While our covariance matrix is approximately stationary (see Appendix \ref{sec:station}), we do not constrain it to be. We also do not require the covariance obey a functional form such that the covariance matrix is positive-definite for all sample points, a common constraint in GPR. More simply, this regression can be viewed as a linear least-squares regression using the known pixel ($k$) values as the regressors to predict the unknown pixel ($k_\star$) values in the high-dimensional pixel space (see notational correspondence in Appendix \ref{sec:LS}). The local shifted samples are then the training data from which the parameters connecting the $k$ and $k_\star$ pixels are estimated. Note that this model is \textbf{not} linear in the 2D spatial dimensions of the image. Even though we do not enforce stationarity, we find it convenient to use the language of GPR throughout.

We first estimate a local covariance matrix ($N_p^2 \times N_p^2$) for a subimage ($N_p \times N_p$ pixels) of the residuals centered on each star. We focus on applying LPI to the residuals because we want to predict non-zero mean deviations between the model and true background, assuming the star was not present in the image (Equation \ref{eq:resLikeli}). It is useful to decompose the $N_p^2$ pixels of the subimage into three classes: ``good,'' which we believe to be uncontaminated samples of the local background, ``hidden,'' which are obscured by the star of interest and should be interpolated, and ``ignored,'' which are contaminated either by other stars, cosmic rays, pixel defects, etc. The ``ignored'' pixels should be excluded from background estimation but need not be interpolated. See Sections \ref{sec:SynVal} and \ref{sec:Survey} for detailed examples of masking in the context of applications.

The goal of LPI is to predict the value of missing data in a vector (i.e. the ``hidden'' pixels) given the data present (i.e. the ``good'' pixels) and a model for the covariance of the data vector. Consider a data vector $X$, of length equal to the number of pixels in a subimage around a star $N_p^2$, and the corresponding mean vector $\mu$ and covariance matrix $C$. We denote the indices in $X$ which are ``good'' $k$ and those that are ``hidden'' $k_\star$. Then $\bar{X}(k_\star|k)$, the mean of the missing values predicted conditionally using the known values, is given by Equation \ref{eq:kpred} and the covariance of the prediction is given by Equation \ref{eq:covpred} \citep{rasmussen2003gaussian}.

\begin{ceqn}
\begin{align} \label{eq:kpred}
\bar{X}(k_\star|k) \equiv \bar{X}'_{k_\star} = C_{k_\star,k}C_{k,k}^{-1}[X_k-\mu_k] + \mu_{k_\star}
\end{align}
\end{ceqn}

Here $C_{k_\star,k}$ denotes the $C$ sub-block of rows with $k_\star$-indices and columns with $k$-indices.

\begin{ceqn}
\begin{align} \label{eq:covpred}
{\rm cov}(X'_{k_\star}-\bar{X}'_{k_\star}) = C_{k_\star,k_\star}-C_{k_\star,k}C_{k,k}^{-1}C_{k,k_\star}
\end{align}
\end{ceqn}

In GPR, $C$ is often specified by a choice of parametric kernel (either chosen or with parameters learned from data). However, in LPI we learn $\hat C$ directly from the data without any parameterization, as described next.

\subsection{Covariance Construction} \label{sec:CovCon}

We learn $\hat C$ (an $N_p^2 \times N_p^2$ matrix) for a subimage of $N_p \times N_p$ pixels (which we can view as a vector $X$ of $N_p^2 \times 1$ numbers). To obtain samples of $X$, we shift our $N_p \times N_p$ image up to $N^{\rm{sample}}_x$ and $N^{\rm{sample}}_y$ pixels horizontally and vertically from the stellar center. Assuming that all of the $N_p \times N_p$ subimages derived from a local region are drawn from the same distribution, we can use these samples to estimate $C$ by $\hat C = V^TV/(N-1) - \mu_X \mu^T_X$ where the rows of $V$ are each a sample of $N_p^2$ pixels, $N$ is the total number of samples, and $\mu^T_X$ is obtained by averaging over rows (samples). For sufficiently small spatial scales, the statistics of the dust/gas should be uniform and this stationarity assumption thus valid.

In practice, constructing $V$ and computing $V^TV$ per star is too computationally expensive. However, given that all of our samples are related simply by pixel shifts of the same image, we can perform the averaging over samples as a spatial average using a computationally inexpensive boxcar mean. Consider an analogous problem in 1D, with samples $V_{ij} = X_{i+(j-1)}$, where $X$ is 1-indexed.

\begin{align} \label{eq:boxcar}
\hat C_{ij} &= \frac{1}{N-1}\sum_{k=1}^{N} V_{ki}V_{kj} - \frac{1}{N^2}\left(\sum_{k=1}^{N} V_{ki}\right) \!\!\! \left(\sum_{k=1}^{N} V_{kj}\right) \nonumber \\ 
&=  \frac{1}{N-1}\sum_{k=1}^{N} X_{k+(i-1)}X_{k+(j-1)} \\
&\quad\quad- \frac{1}{N^2}\left(\sum_{k=1}^{N} X_{k+(i-1)}\right) \!\!\! \left(\sum_{k=1}^{N} X_{k+(j-1)}\right) \nonumber
\end{align}

The elements of $V^TV$ are simply the elements of a boxcar mean of width $N$ on the product of $X$ with itself shifted by $(i-j)$. This can be quickly computed for the entire image, and then referenced at the location of each star. Since we consider 2D image data, we implement Equation \ref{eq:boxcar} with an additional spatial index. For the application in Section \ref{sec:Survey}, this reordering of the calculation is four orders of magnitude faster than a direct matrix multiplication per star for fields with the median number of stars (see Section \ref{sec:CompCost}). 

\subsection{Preliminary Infilling} \label{sec:prelim_infill}

The local covariance estimate must be computed over ``good'' pixels, excluding ``hidden'' and ``ignored'' pixels. Unfortunately the naive approach of considering the masked, shifted boxcar means fails because summing over only masked pixels in Equation \ref{eq:boxcar} does not correspond to $V^TV$, and the resulting matrix is no longer guaranteed to be positive semidefinite \citep{Baghi:2016:PhRvD:}. Instead, we must make a reasonable guess to infill the masked pixels. As long as the masking fraction is low, the covariance matrix estimate is not strongly sensitive to the infill.

Our infilling procedure uses a boxcar mean of the ``good'' pixels to predict masked pixels using a variety of filter widths (both “hidden” and “ignored” pixels must be infilled in this initial step). The initial width is 19 pixels, and all pixels with a mean value determined by 10+ good pixels are infilled. The filter width is increased by a factor of 1.4 in successive infilling steps until all of the masked pixels are infilled. In heavily crowded/masked regions, it is useful to have a cut-off at a finite number of infilling steps (10), and infill the remaining pixels with the median image value. These numerical choices were qualitatively chosen for performance on the application in Section \ref{sec:Survey}.

These infilled pixels should also have per-pixel uncertainties consistent with the rest of the image, so we replace each masked pixel by a Poisson draw. However, we must account for the counts already subtracted out in the smooth background model and detector gain, which pools $g$ photoelectron counts per digital unit.\footnote{The background plus residuals should never be negative as such negativity would indicate that the model image included sources with negative flux or nonpositive portions of the PSF. Unfortunately, both of these situations can occur in \texttt{crowdsource} outputs on difficult fields, so we take $|\rm{background + residuals}|$ in order to report a value that the user can decide to use based on quality flags.}

\begin{ceqn}
\begin{align} \label{eq:infill_draws}
g \times (\hat{b} + \hat{r}) \sim {\operatorname {Poisson}}(g \times (\hat{b} + \bar{r}))
\end{align}
\end{ceqn}

\subsection{Deriving Corrections} \label{sec:Correct}

One of the benefits of the LPI background model is the ease of deriving improved flux estimates. The maximum likelihood estimate for the flux correction to a star can be found by setting the first derivative of the log-likelihood $\ln
\Lagr(f^{\rm{b}})$ over the $k_\star$ pixels with respect to the bias offset $f^{\rm{b}}$ equal to zero. Similarly, the variance of the new flux estimate $\hat{f} - f^{\rm{b}}$ is obtained as the negative inverse of the second derivative of the log-likelihood with respect to the flux estimate.

\begin{equation}
\begin{aligned} \label{eq:MLvar}
-\ln \Lagr \propto
(f^{\rm{b}} \ p + r - \bar{X}'_{k_\star})^T \hat C^{-1} (f^{\rm{b}} \ p + r - \bar{X}'_{k_\star})
\end{aligned}
\end{equation}

\begin{ceqn}
\begin{align} \label{eq:debiasFactors}
f^{\rm{b}} = - \frac{p^T \hat C^{-1}r}{p^T\hat C^{-1}p} + \frac{p^T\hat C^{-1}\bar{X}'_{k_\star}}{p^T \hat C^{-1}p} = f^{\rm{b}}_{\rm{res}} - f^{\rm{b}}_{\rm{pred}}
\end{align}
\end{ceqn}

\begin{ceqn}
\begin{align} \label{eq:debiasError}
\sigma^2_{f+f^{\rm{b}}} = -(\partial^2_{f+f^{\rm{b}}} \ln \Lagr)^{-1} = (p^T \hat C^{-1}p)^{-1}
\end{align}
\end{ceqn}
We have suppressed the per-star PSF ($p$) position evaluation and $i$ subscripts for clarity, highlighting the $f^{\rm{b}}$ dependence. Because $\hat C$ learned in Section \ref{sec:CovCon} was only estimated on pixels representative of uncontaminated background, we must adjust the per-pixel variance (diagonal) of $\hat C$ to account for (Poisson) noise from modeled counts before computing Equations \ref{eq:debiasFactors} and \ref{eq:debiasError} (see Section \ref{sec:decamjl}).

We interpret the first term, which we denote $f^{\rm{b}}_{\rm{res}}$, as accounting for the fact that the initial photometric pipeline might not be able to achieve its target of zero residuals through modeling point sources. This often leads to overestimation of the flux from stars to soak up counts from bright structured backgrounds. In the worst case, photometric pipelines can add a large number of nonexistent sources to the model in order to account for the background flux. This suppresses evidence of background mismodeling in the residuals and is a failure mode beyond the scope of this work. 

The second term, which we denote $-f^{\rm{b}}_{\rm{pred}}$, measures the expected deviation from zero residuals with respect to a flat background model as a result of the interpolated structured background. A pictorial scheme illustrating the role of these terms is depicted in Figure \ref{fig:correctFactScheme}. We packaged all of the above steps required to compute LPI corrections to photometry into a code, \textsc{CloudCovErr.jl} which we release with this work (see Section \ref{sec:dataavil}). 

An alternative approach to the one pursued here would be to incorporate the predicted background and non-diagonal covariance matrix into Equation \ref{eq:resLikeli} as an improved $\hat b$ and $\hat C$. In this way, the improved background model and uncertainties are incorporated into the original photometric solution. One difficulty with this approach is that we currently require an initial photometric solution in order to have access to residuals. Another complication is that the implementation will depend on whether the photometric pipeline solves for the flux of each source independently, or as a joint optimization problem. For these reasons, we present a pure afterburner as a proof-of-concept, which is conveniently compatible with any prior photometric solution.

\section{Gaussian Approximation Validity} \label{sec:Approx}

\begin{figure}[b]
\centering
\includegraphics[width=\linewidth]{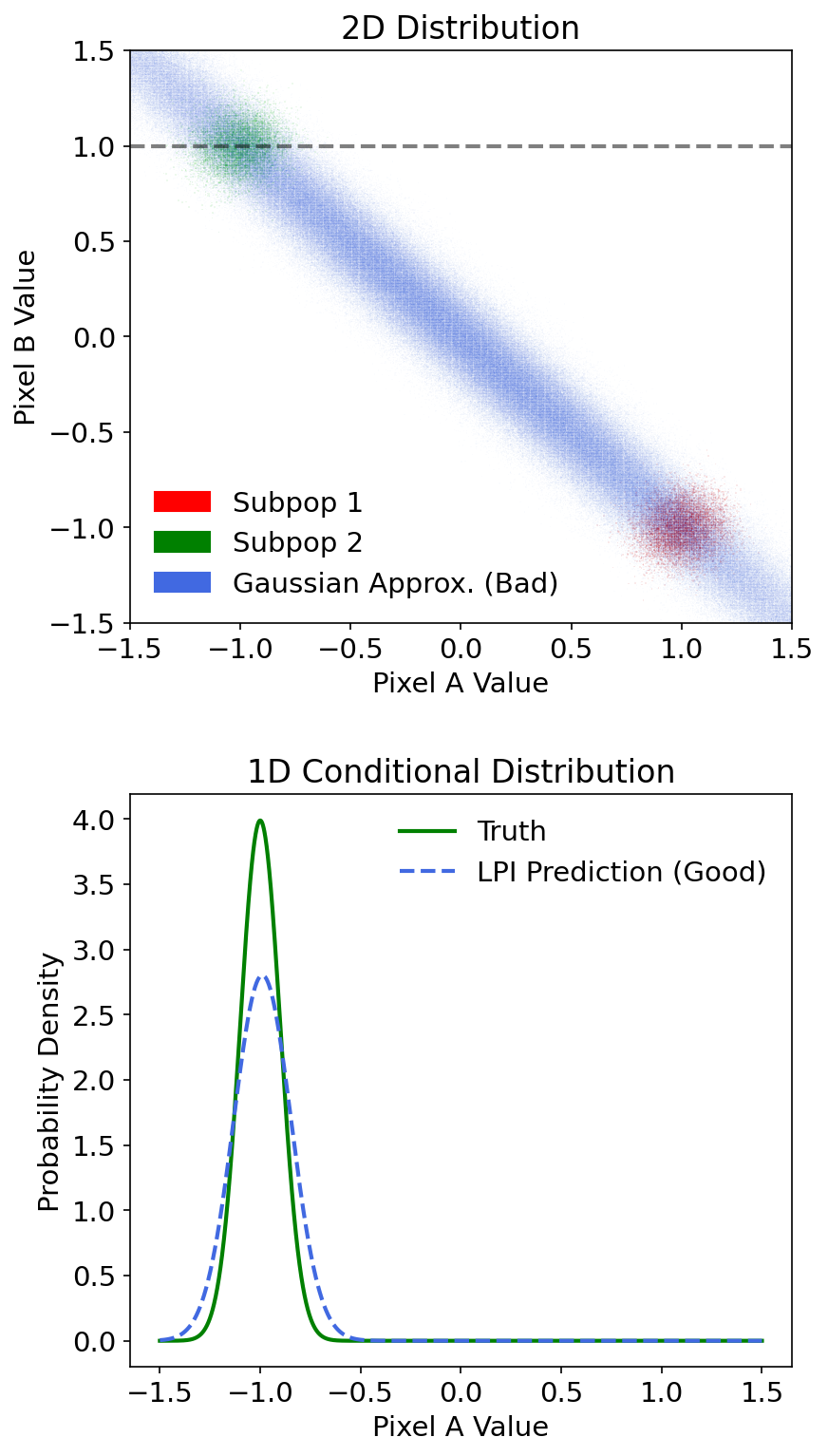}
\caption{\textit{Top}: Probability distribution for population in toy 2D problem. The two Gaussian subpopulations are colored red and green. The Gaussian approximation to the distribution is shown in blue, and poorly describes the total population. A line cut along the grey dashed line yields the bottom panel. \textit{Bottom}: The conditional probability distribution for pixel A given pixel B has a value of 1.0. While the true distribution (green) is more sharply peaked, the Gaussian process regression prediction (blue-dashed) describes the conditional distribution well.}
\label{fig:toy}
\end{figure}

\begin{figure*}[t]
\centering
\includegraphics[width=\linewidth]{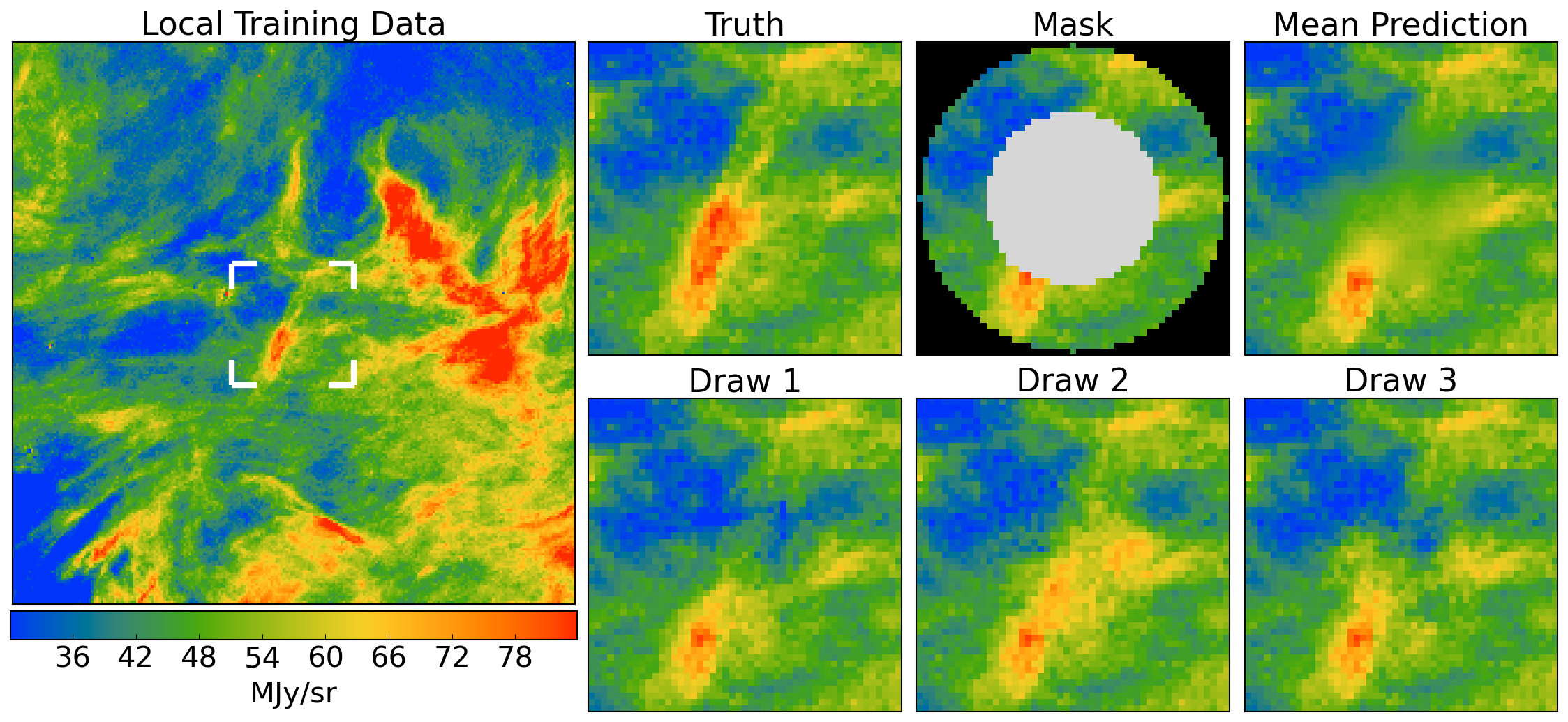}
\caption{\textit{Left}: Image of dust emission from WISE 12 $\mu$m dust map containing a portion of the Pipe Nebula. The image scale shows all data used to train the LPI covariance, which is highly heterogeneous. Square marked by four white corners indicates subimage used in other panels. \textit{Top Right}: ``Truth'' shows the central $N_p = 49$ pixel square cut-out of the WISE image. In ``Mask'', the central grey pixels are those predicted (``hidden'') and the outer black pixels not used (``ignored'') during the prediction. In ``Mean Prediction'', the ``hidden'' pixels are replaced with the mean LPI predictions and all other pixels display the true values. \textit{Bottom Right}: Each image replaces the ``hidden'' pixels with a different random draw from the distribution of LPI predictions. All other pixels display the true values. Videos animating this plot as the size of the ``ignored'' region (\href{https://faun.rc.fas.harvard.edu/saydjari/CloudCovErr/rseries_test.mp4}{\textbf{link}}) and ``hidden'' region (\href{https://faun.rc.fas.harvard.edu/saydjari/CloudCovErr/thr_test.mp4}{\textbf{link}}) is changed demonstrate that these interpolations are robust, even for a narrow ring of ``good'' pixels.} 
\label{fig:WISE}
\end{figure*}

In the ISM, there are isolated dust clouds, filaments of gas, and shock fronts all of which are highly non-Gaussian features. This is in contrast to cosmic microwave background cosmology where the structures of interest are well-described by Gaussian random fields and small deviations from Gaussianity are of interest. As a result of this strong heterogeneity, the non-Gaussian structure of the ISM cannot even be described as some functional transform of a Gaussian random field, such as by a log-normal random field. In order to draw the correspondence to Gaussian process methods and evaluate the validity of using LPI in the context of these non-Gaussian fields, one can view the non-Gaussian field as being represented as a Gaussian mixture model-- a sum of many different Gaussian components.

Naively, one might expect LPI to be valid only when the local pixels are drawn from a single component of the Gaussian mixture model and the local image is thus locally Gaussian. We will show in the context of a toy problem that LPI works well in a larger class of cases where we only require that the conditional distribution for the $k_{\star}$ pixels given the known $k$ pixels is Gaussian. Then, we will demonstrate that the LPI works well in practice on a real, filamentary dust image.

\subsection{Toy Problem} \label{sec:Toy}

Consider an image consisting of two pixels, pixel A and pixel B. Suppose the image is drawn from a population with a $50\%$ chance of being drawn from $\mathcal{N}(\mu=[-1,1],\sigma=[\epsilon,\epsilon])$ or $\mathcal{N}(\mu=[1,-1],\sigma=[\epsilon,\epsilon])$. The distribution of the pixels in this subimage is clearly non-Gaussian, composed of two separated Gaussian components (Figure \ref{fig:toy}). Further, the Gaussian approximation to this distribution is quite poor, placing maximal probability on the infinitesimally rare image [0,0].

However, when one pixel value is known (say pixel B without any loss of generality), the LPI prediction provides a good approximation of the distribution of the unknown pixel. Pictorially, the predicted conditional distribution on pixel A given the known value of pixel B is simply a renormalized horizontal line cut through Figure \ref{fig:toy} at that value of pixel B. Numerically, we follow Equations \ref{eq:kpred} and \ref{eq:covpred} to obtain $\mathcal{N}(\mu=-1/(1+\epsilon^2),\sigma=\sqrt{2}\epsilon)$, a close approximation to $\mathcal{N}(\mu=-1,\sigma=\epsilon)$ for pixel A when pixel B $= 1$, though with inflated variance.

The success of this model problem lends confidence to LPI being able to correctly interpolate non-Gaussian local subfields that may contain filaments. Two adjacent pixels transverse to the long filament direction will be highly anti-correlated on the positive or negative slope off of the peak of the filament. This will lead to two Gaussian components similar to those in Figure \ref{fig:toy}. When considering the slope changing along the transverse direction, more pairs of Gaussian components closer or farther from the origin will be added to the population. However, a GMM of components along the diagonal are all clearly still resolvable by LPI. It is only when there are components of the GMM which are not distinguished by the ``good'' conditioning pixels that LPI can be a poor approximation.

\subsection{Interpolating in a Nebula} \label{sec:WISE}

To demonstrate that the assumptions of LPI are compatible in practice with astronomical fields containing structured backgrounds, we mask and interpolate a real image from the WISE 12 $\mu m$ dust map \citep{Meisner:2014:ApJ:} corresponding to a portion of the Pipe Nebula (between Barnard 67 and 78) (Figure \ref{fig:WISE}). We preprocess the initial image by first clamping outliers to the highest and lowest $0.01\%$-percentile and then reducing the resolution from 6" to 45"/pixel.\footnote{The WISE 12 $\mu m$ dust map was smoothed during production to 15" so this is only a 3x reduction in resolution. We downsample to 45"/pixel only to put the features of interest at a pixel scale which is computationally inexpensive.} The covariance matrix for a $N_p = 49$ pixel subimage was estimated from samples on a larger local region (using a boxcar width of $N_x^{\rm{sample}}=N_y^{\rm{sample}}=129$ pixels), shown in the left panel of Figure \ref{fig:WISE}. 

The true $N_p \times N_p$ subimage is shown in the ``Truth'' subpanel and features a core with filamentary connections toward dense regions at the top right and right of the subimage. In the masked image, the central mask with radius $\sim$13 pixels defines the $k_\star$-pixels (``hidden''). The known true values of the ``good'' $k$-pixels are shown. The pixels far away from the interpolation region are ``ignored'' since they should have little impact on the interpolation and add to the computational expense. Videos animating Figure \ref{fig:WISE} as the sizes of the ``ignored'' region (\href{https://faun.rc.fas.harvard.edu/saydjari/CloudCovErr/rseries_test.mp4}{\textbf{link}}) and ``hidden'' region (\href{https://faun.rc.fas.harvard.edu/saydjari/CloudCovErr/thr_test.mp4}{\textbf{link}}) are changed demonstrate that the interpolation is robust to these choices.

\begin{figure}[b]
\centering
\includegraphics[width=\linewidth]{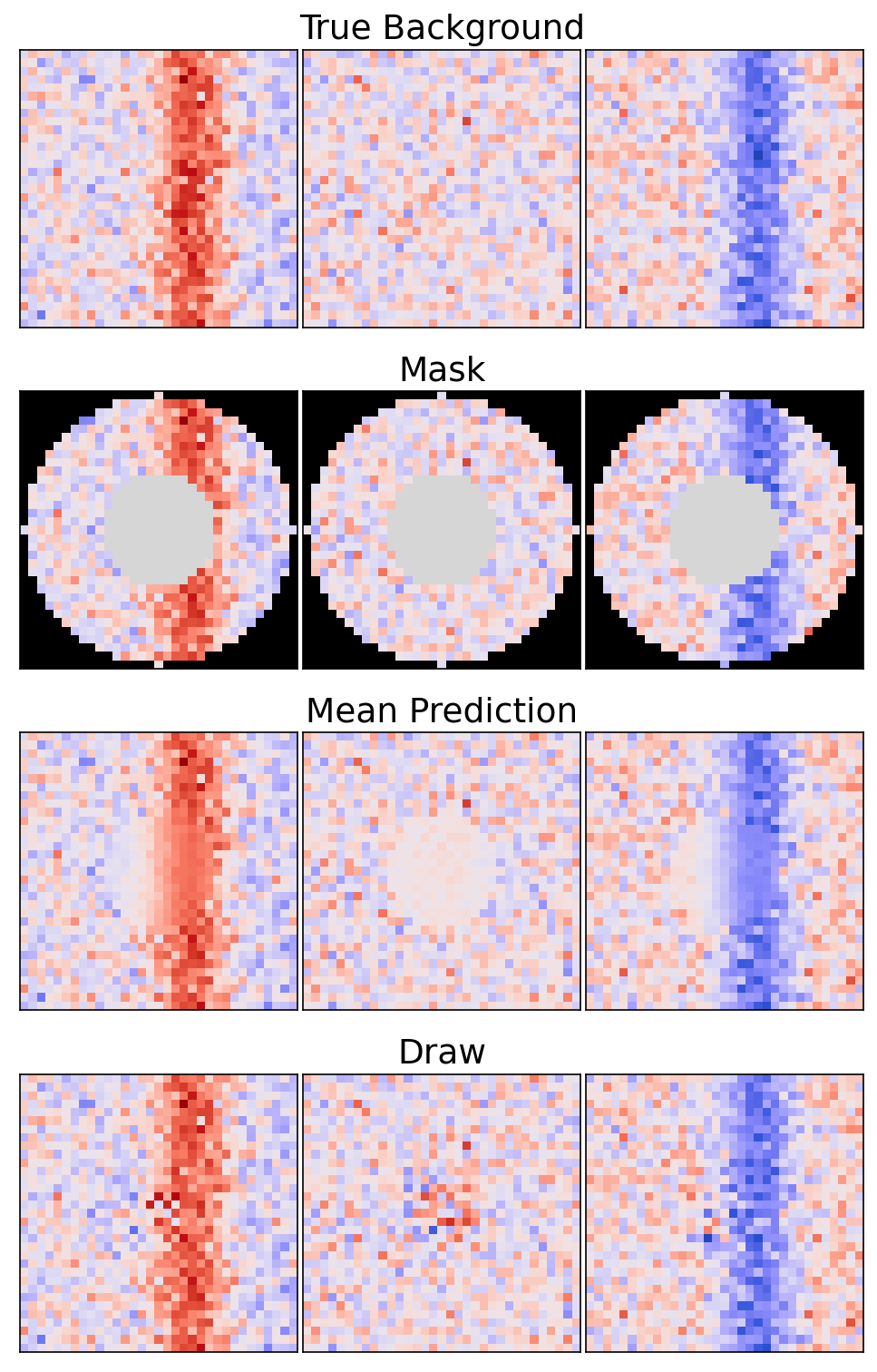}
\caption{The columns show a stellar position near a positive filament (\textit{left}), in between the filaments (\textit{middle}), and near a negative filament (\textit{right}) in the residuals of a synthetic image (red positive, blue negative). \textit{Top Row}: The true background before source injection is shown. \textit{Second Row}: The mask of ``hidden'' and ``ignored'' pixels is shown, as in Figure \ref{fig:WISE}. \textit{Third Row}: The ``hidden'' pixels are replaced with the mean local pixelwise infilling (LPI) prediction. \textit{Bottom Row}: The ``hidden'' pixels are replaced with a draw from the LPI distribution.}
\label{fig:FilRecon}
\end{figure}

The mean prediction of the LPI is shown replacing the values of the $k_\star$-pixels in the true image. Comparing this mean prediction to the ``Truth'', we see that near the core, the interpolation correctly predicts a higher density, and on average predicts an overdensity extending toward the dense region at the right of the image. For the diagonal filament, the mean predicted density appears to have a much lower intensity between the two cores than the ``True'' image. However, the mean does not contain all the information provided by the LPI, and we must also consider the covariance.

One way to visualize the covariance of the LPI is to sample the distribution of interpolated images. Samples are easily obtained as $\sqrt{\hat C}y + \bar{X}'_{k_\star}$ where $y \sim \mathcal{N}(0,1)$, $\bar{X}'_{k_\star}$ is the predicted mean, and $\sqrt{\hat C}$ is the symmetric square root, which we obtain via singular value decomposition (SVD). Three different draws are shown in the bottom row of Figure \ref{fig:WISE}. While none of these draws reproduces the true image, the interpolated pixels have noise statistics which match the true image and some of the draws do have filaments connecting the central core to both overdense regions (i.e. Draw 2). Thus, the LPI is consistent with the ``Truth'', even though the LPI is not centered on the ``Truth.''

\section{Synthetic Validation} \label{sec:SynVal}

In order to show that our LPI bias offsets and error bars provide a substantial improvement over current photometric pipelines, we run our post-processing code on outputs from \texttt{crowdsource}, a crowded-field photometric pipeline that was used to create catalogues from DECam and WISE \citep{Schlafly:2018:ApJS:,Schlafly:2019:ApJS:}. First we consider synthetic fields here, and then astrophysical images from the DECaPS2 survey in Section \ref{sec:Survey}.

\subsection{Treatment of Emission and Extinction} \label{sec:Fil}

The synthetic data we consider consists of two filaments, each a Gaussian profile with full width at half max (FWHM) of 6.7 pixels. One filament is positive and one negative relative to the background. The filaments are separated by 118 pixels and the background and Poisson noise are consistent with fields from DECaPS2.\footnote{For this reason, we quote flux in analog-to-digital units (ADU) throughout synthetic tests.} Stars, all of the same flux ($\sim$18th g-band magnitude), are then injected on a regular grid with centers at integer pixel values. Each row in the grid is shifted by 1 pixel in the direction transverse to the filament in order to sample stars with different displacements relative to the filament. We then run \texttt{crowdsource} on the synthetic image. We fix the PSF model to match the model used to inject sources and use less aggressive deblending than defaults to prevent breaking the filament into nonphysical point sources.\footnote{We set a higher threshold for detection (10) and set the ``nodeblend'' and ``sharp'' mask bits. See \cite{Schlafly:2018:ApJS:} for a detailed description of these parameters.}

We post-process the \texttt{crowdsource} outputs using the code base developed in this work, \textsc{CloudCovErr.jl}, with $N_p=33$ and $N^{\rm{sample}}_x=N^{\rm{sample}}_y=129$. The top panel of Figure \ref{fig:FilRecon} shows the residual image for the true background given the \texttt{crowdsource} background model. Each column shows a subimage centered on a star near the positive filament, negative filament, or between filaments. The $k_\star$-pixel mask is obtained by a threshold on the PSF model used for source injection and masks a radius of $\sim 10$ pixels. Pixels beyond a radius of 16 were ``ignored'' (see second row of Figure \ref{fig:FilRecon}). For the third and fourth rows, the $k_\star$-pixels around each star are replaced by the mean prediction or by draws from the predicted distribution.

While the mean predictions reconstruct the filament position within the $k_\star$-pixels correctly, the exact height of the filament peak is often slightly underestimated. We will explore the magnitude of this underestimation in the Section \ref{sec:SNR}. The draw rows show interpolation that is almost indistinguishable from the background except for a slightly higher noise level, suggesting a slight overestimation of the uncertainty. However, that is expected. The averaging region for estimating the covariance matrix here is large and includes both filaments, a more heterogeneous environment than any of the regions reconstructed.

\begin{figure}[b]
\centering
\includegraphics[width=\linewidth]{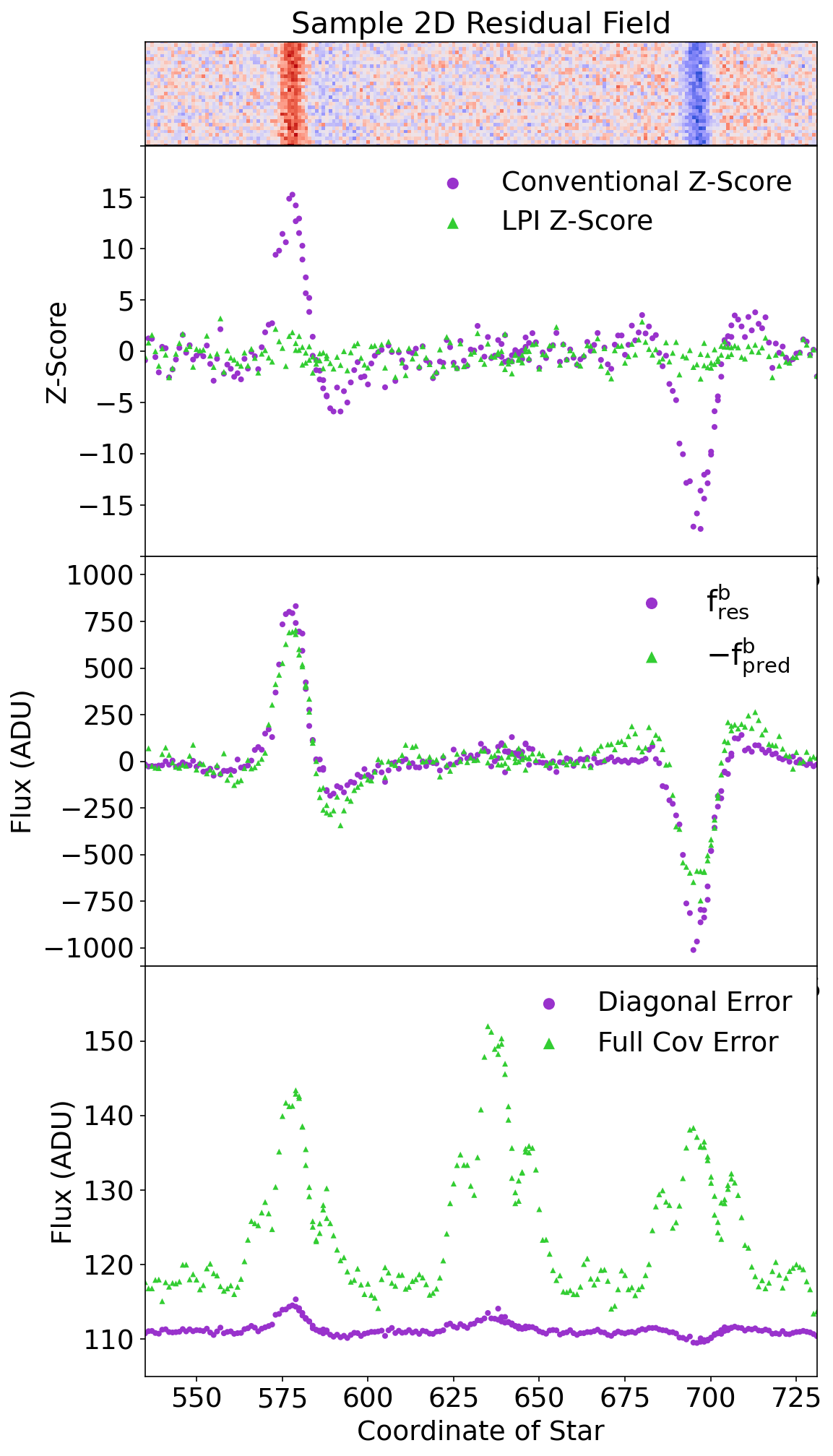}
\caption{\textit{Top}: Residual field (2D) from synthetic image where the x-coordinate matches the 1D stellar coordinate used below. \textit{Second}: Z-scores for injected stars as a function of position using conventional (purple circle) or LPI photometry (green triangle). The absence of positive and negative Z-scores correlated with the filament demonstrates that LPI corrects for the structured background. \textit{Third}: The two debiasing terms, $\rm{f}^{\rm{b}}_{\rm{res}}$ (purple circle) and $-\rm{f}^{\rm{b}}_{\rm{pred}}$ (green triangle), are shown to contribute similarly to correcting for the filaments. Flux is in analog-to-digital units (ADU). \textit{Bottom}: The predicted flux uncertainty is shown, assuming the covariance matrix is diagonal (purple circle) or using the full covariance matrix (green triangle).}\vspace{-4mm}
\label{fig:FilTest}
\end{figure}

The second panel of Figure \ref{fig:FilTest} shows that the Z-scores (observed flux - true flux)/(flux uncertainty) for the \texttt{crowdsource} solutions (purple circles) can be as large as $\pm 15 \sigma$ and are correlated with the filament position. Note that these conventional Z-scores use uncertainties assuming the covariance matrix is diagonal, only accounting for the uncorrelated Poisson noise. In contrast, the Z-scores that include the bias offset we derive here and use error bars accounting for non-diagonal covariance (green triangles) are flat across the filament positions and are consistent with a standard normal distribution. In the third panel of Figure \ref{fig:FilTest}, we see that the two bias offsets defined in Equation \ref{eq:debiasFactors} have approximately equal contribution in this case.

One method of quantifying the normality of the Z-score distribution is evaluating the cumulative distribution function (CDF). However, when only a small fraction of the stars in the population are affected, a large local improvement such as the one shown in Figure \ref{fig:FilTest} can appear as only a slight improvement in the CDF. In fact, one of the most common ways of ``fixing'' error bars in astronomy is to modify the error bars until the CDF matches a normal distribution. The modifications include adding an uncertainty floor, adding a fractional uncertainty (generally ascribed to absolute calibration errors), or simply rescaling the error bars by a scalar \citep{Fitzpatrick:2009:ApJ:}. While these modifications improve the distribution of the entire population, they ignore the detailed local environment of each star, in contrast to LPI photometry. Further, all of those previous distributional approaches will increase the uncertainty estimates for emission or reflection (positive filament) more than for extinction (negative filament).

In the fourth panel of Figure \ref{fig:FilTest}, we plot the predicted flux uncertainty assuming the covariance matrix is diagonal (purple circles) and using the full covariance matrix (including off-diagonal contributions, green triangles). For the diagonal error bars, there is a small peak for the positive filament and a small dip for the negative filament which follows the independent Poisson pixel measurement noise. Thus, no method of rescaling the diagonal error bars can symmetrically treat both the positive and negative filament. 

\begin{figure}[b]
\centering
\includegraphics[width=0.93\linewidth]{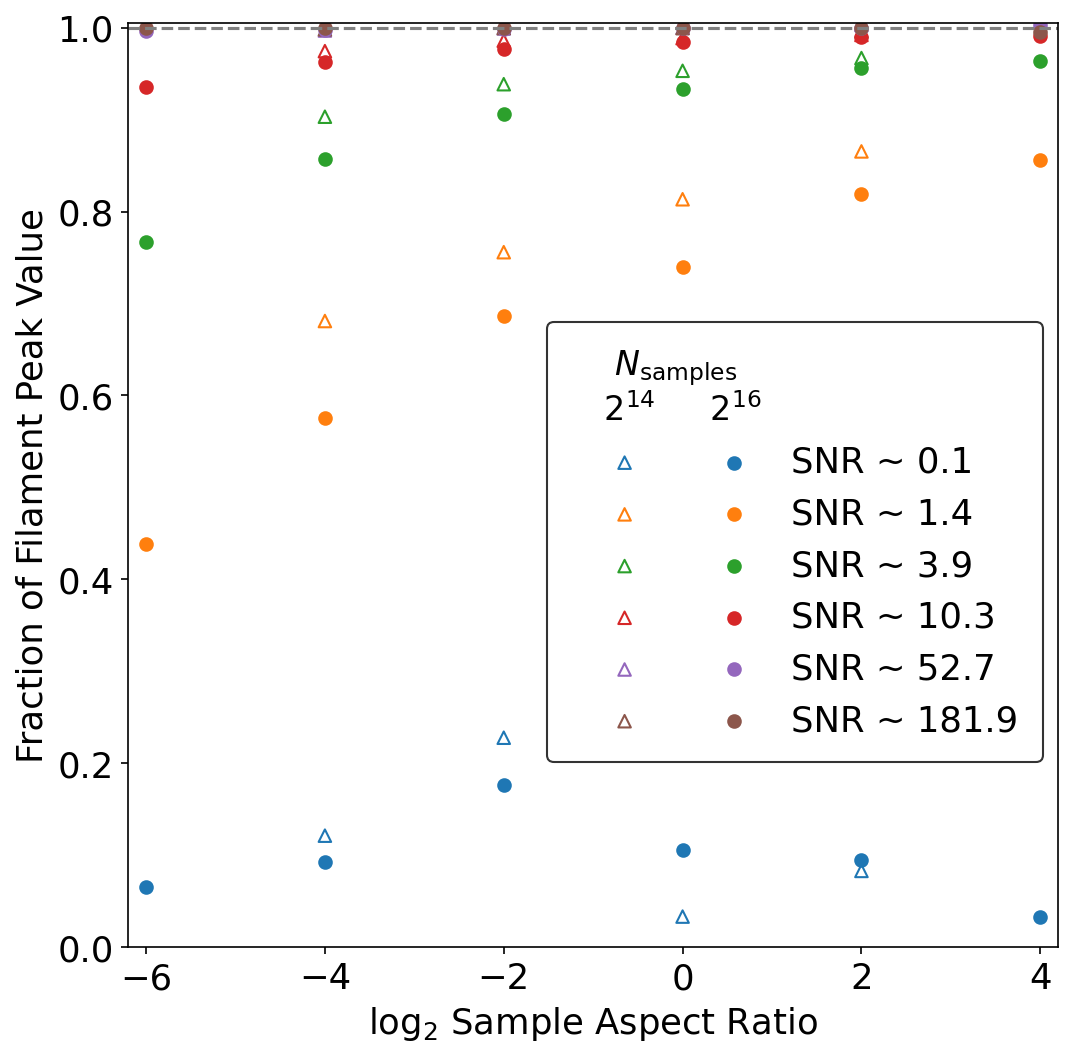}
\caption{Convergence of the predicted filament height as a function of signal-to-noise ratio (SNR). To the right along the horizontal axis, the fraction of samples (used for estimating the covariance matrix) along the filament increases. The total number of samples is fixed to either $2^{14}$ (open triangle) or $2^{16}$ (closed circle) by changing the total area of the boxcar window. The main effect on convergence of increasing sample size is the additional samples transverse to the filament, causing an apparent shift of $+2$ along the aspect ratio axis with respect to the triangle markers. For SNR $\geq 1$, a large fraction of the filament height is reconstructed.}\vspace{-2.7mm}
\label{fig:SNR}
\end{figure}

\begin{figure*}[t]
\centering
\includegraphics[width=\linewidth]{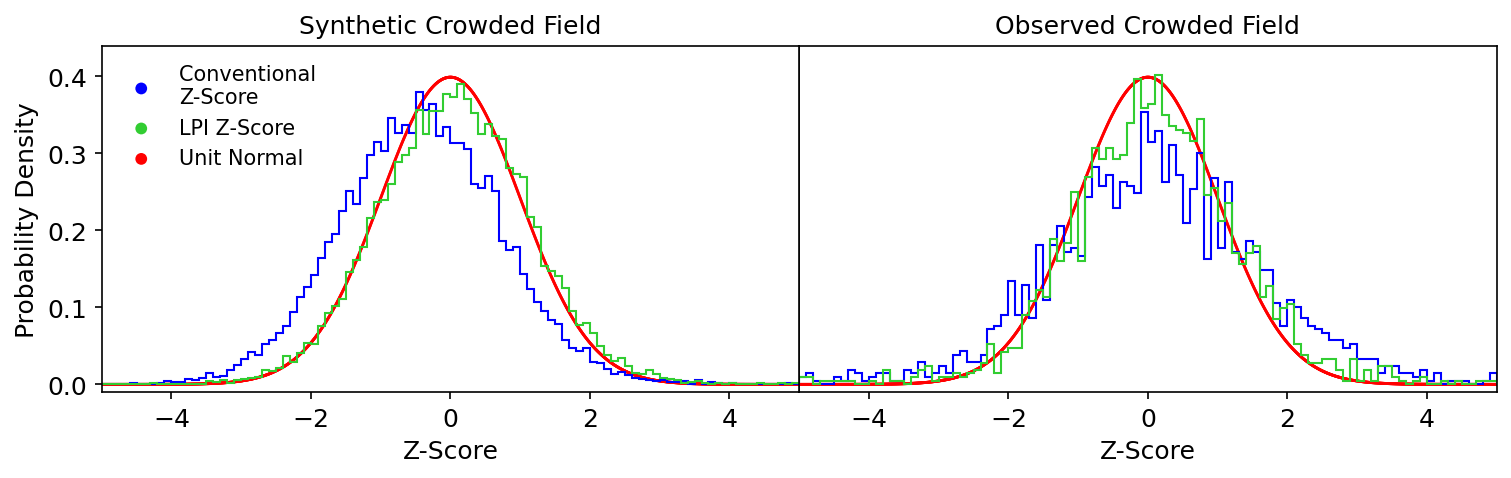}
\caption{Probability distribution function (PDF) of the conventional Z-scores (blue), LPI Z-scores (green), and a unit normal distribution (red), for reference. \textit{Left}: There is a clear leftward shift in the conventional PDF, indicative of flux underestimation, which is not present in the LPI PDF. Thus, the major improvement of LPI photometry in the synthetic crowded field is correcting bias in the background model. \textit{Right}: The major improvement for the real observed field is increasing the reported flux uncertainty. This manifests as the LPI PDF being more peaked and closer to the normal distribution compared to the conventional PDF (which is too broad). In both cases, the improved match between the green and red curves indicates the improved performance of LPI photometry.}
\label{fig:crowdedflat}
\end{figure*}

In contrast, the error bars derived from the full covariance matrix peak over both the positive and negative filaments. Away from the filaments on flat sky, the full error bars approach the diagonal approximation, agreeing with the intuition that away from coherent structure, the uncertainty is dominated by the independent per-pixel measurement noise. For stars on flat sky in between the filament, the covariance matrix estimate is trained on the local region containing both filaments. The predicted uncertainty then displays large off-diagonal contributions. This is of course a limitation of the local covariance model; it can be influenced by local structure, even if that structure is not affecting the star of interest. This limitation drives us to make the number of samples (the size of the boxcar windows) as small as possible while maintaining numerical stability for matrix inversion.

\subsection{Role of SNR and Aspect Ratio} \label{sec:SNR}

It is important to understand how the quality of the LPI of the filament is a function of the SNR of the data, local environment used to estimate the covariance matrix, and number of training samples. To do this, we mask and interpolate 34 equally-spaced positions centered along a synthetic filament with a masking radius of 5 pixels and $N_p = 33$. We report the average of the row of (11) pixels along the filament spine, pooling over all 34 reconstructions, as a fraction of the true filament height (Figure \ref{fig:SNR}). 

For what we call a sample aspect ratio ($\log_2$) of $0$, we learn the covariance matrix over a symmetric region, using a boxcar width of $129$ pixels in both the $x$ and $y$ directions ($N^{\rm{sample}}_xN^{\rm{sample}}_y = 2^{14}$) and obtain a reconstruction of $>80\%$ of the true peak value at an SNR $\sim 1.4$. As SNR increases, the predicted peak height converges to the true value, further validating the application of the LPI method to non-Gaussian structures. In the low SNR limit (SNR $\sim 0.1$), the noise dominates the reconstruction and there is only a small positive bias due to the filament, as expected.

Instead of using a symmetric sampling of the $x$ and $y$ directions, we can consider more samples along the filament direction (positive sample aspect ratio) or transverse to the filament direction (negative sample aspect ratio). This amounts to using a boxcar width of (255,65) for sample aspect ratio $2$, given the filament along the $x$ direction. The reconstruction converges more quickly the more samples are along the filament direction as opposed to the transverse direction. This is consistent with the intuition that in order to reconstruct the filament, the conditioning must overcome a large bias in the model for a flat background when a large fraction of the training sample is flat (as it is in the transverse direction relative to the filament). For numerical stability we must truncate the aspect ratios in the positive direction before there are $<2N_p$ transverse samples because the image is translationally invariant (other than Poisson noise) along the filament direction. Increasing the overall number of samples (circle markers, Figure \ref{fig:SNR}) improves numerical stability, but the main effect on convergence is the additional samples transverse to the filament (an apparent shift of $+2$ along the aspect ratio axis with respect to the triangle markers).

\subsection{Toward the Crowded Field Limit} \label{sec:crowded}

To probe behavior of LPI photometry in the crowded field limit, we inject stars, all of the same flux ($\sim$18th g-band magnitude), onto a flat sky background consistent with fields from DECaPS2. We inject 25,000 sources at random integer positions (excluding a buffer around the edge) into a single 2k x 4k pixel CCD image. This source density is 10$\times$ larger than that required to have isolated PSF footprints (given the $59 \times 59$ pixel footprint used by \texttt{crowdsource} for stars of this magnitude). We use only one magnitude of star (and with large SNR) to maximize the probability of close sources being deblended.

Because there is a minimum separation below which two sources cannot be deblended, there is a population of detections that model two close stars as a single star with approximately twice the flux. The larger residuals at the source location resulting from this mismodeling are masked from the LPI covariance matrix estimation, and no large perturbation of the (``good'') $k$-pixels will indicate that a second star is acting as a ``background'' to the other. In other words, LPI works only on structured local backgrounds which are not entirely masked by the star. Thus both conventional and LPI Z-score for such sources are large and positive. To cleanly remove merged detections, we exclude stars with separation less than 4/3 FWHM in the tests that follow ($\sim 6$ pixels).

We display the PDF for sufficiently isolated detections in Figure \ref{fig:crowdedflat}. The PDF for the LPI Z-scores is centered at $\sim 0$, while the conventional Z-scores are left-shifted. This reflects one of the main biases in \texttt{crowdsource}; incompletely deblending sources from the sky background leads to flux underestimation, especially in crowded fields.\footnote{In contrast, other algorithms such as DAOPHOT were found to be biased high by this ``confusion'' noise \citep{Ferrarese:2000:PASP:}} The main benefit of the LPI Z-scores in this case is to provide a better sky estimation. The simply translational difference between the PDFs agrees with the intuition that there is little off-diagonal contribution to the flux uncertainties in this synthetic case.

To make a comparison with a real image, we use a ``crowded'' field g-band image from DECaPS2 (see Section \ref{sec:Survey}) with $\sim 73,000$ detected sources (Figure \ref{fig:crowdedflat}, right). We inject 10\% more synthetic sources using the same PSF model obtained by \texttt{crowdsource} for the original image, sampled from the flux distribution of sources in the original image, excluding sources below a minimum separation of 4/3 FWHM. In this case, both the conventional and LPI PDFs are well centered at $\sim 0$, but the conventional PDF has much larger tails. Thus, the main benefit of the LPI Z-scores on this real field is providing better uncertainty estimates. Compared to the synthetic field, the real field contains more stars ($73,000 > 25,000$), but only a small fraction of them are $18^{\rm{th}}$-magnitude or brighter (387) and we expect the leakage of flux from bright sources to more significantly affect the sky estimation. Therefore, both in the limit of poor sky estimation or correlated residuals from mismodeling, the LPI Z-scores improve the overall PDF in the case of crowded fields.

\section{Validation on a Large Survey} \label{sec:Survey}

\begin{figure}[b]
\centering
\includegraphics[width=0.99\linewidth]{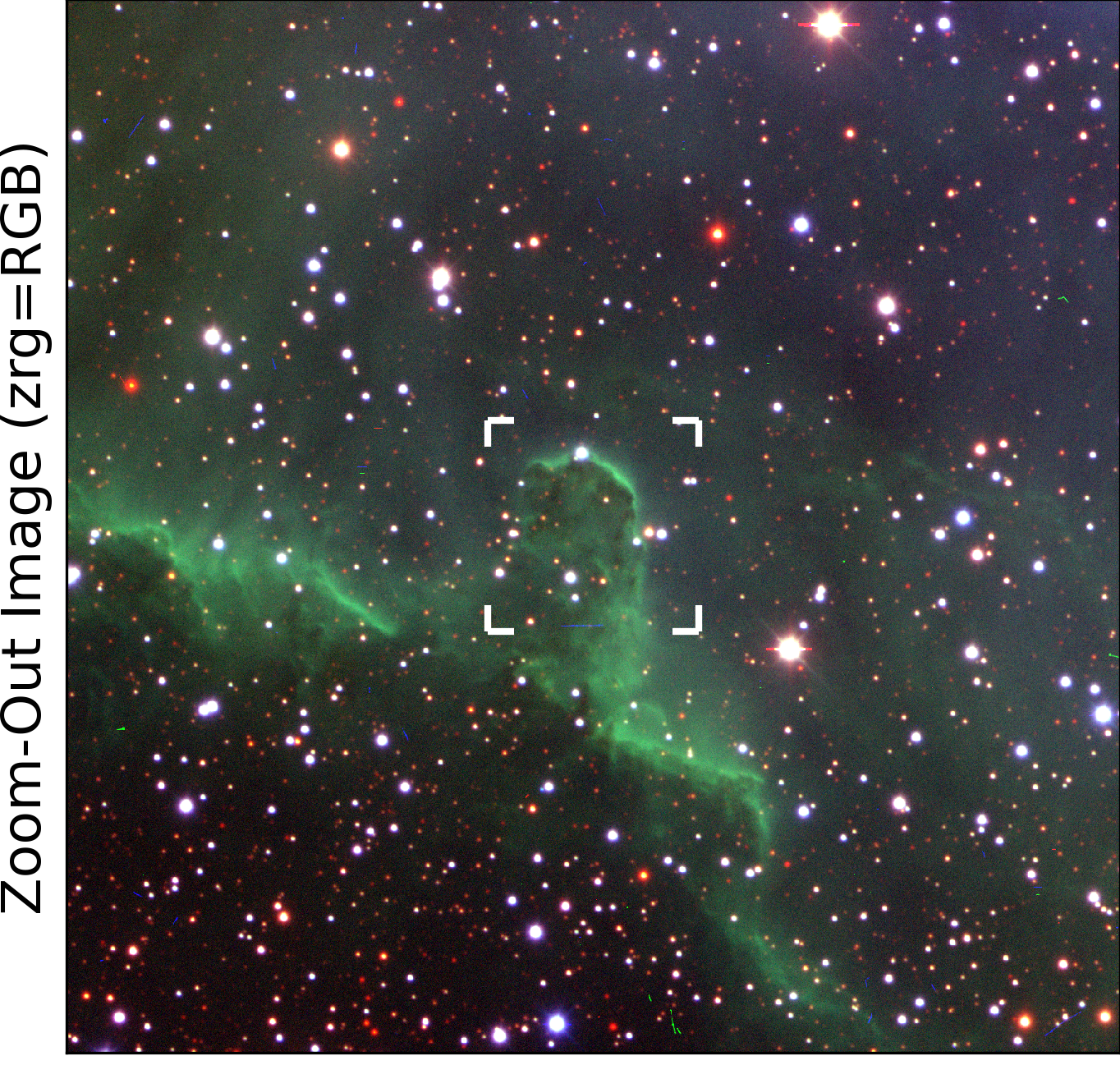}
\caption{Color image (RGB=zrg bands) of larger region where H$\alpha$ emission complicates photometry. Square marked by four white corners indicates subimage used in Figure \ref{fig:CED116}.}
\label{fig:CED116im}
\end{figure}

We now turn to bias offsets and error bars computed on astronomical images from the Dark Energy Camera (DECam) as part of the second release of the Dark Energy Camera Plane Survey (DECaPS2) \citep{Saydjari:2022:ip:}. Briefly, DECaPS2 imaged $|b|<10\degree$, $5\degree > \ell > -120\degree$ in $grizY$ filters, finding 34 billion detections (3.3 billion objects). We use DECaPS2 to demonstrate that underestimated error bars correlated with structure are present in most modern surveys of fields with structured backgrounds, our method is robust to the details of real data, and our method is scalable enough to run on all 34 billion detections from one of the largest photometric catalogs to date.\footnote{DECaPS2 with 3.31B sources is comparable in size to Pan-STARRS1 with 2.9B sources \citep{magnier2020pan}, though DECaPS2 has fewer epochs and thus fewer detections. The NOAO source catalogue (NSC) \citep{nidever2018first}, which used SExtractor to uniformly reprocess public data including DECaPS1, contains 2.9 B. The Zwicky Transient Facility (Data Release 8) reports $>3.2$ B objects \citep{Masci:2019:PASP:}.}

\subsection{CCD Level Processing} \label{sec:decamjl}

Within \textsc{CloudCovErr.jl}, we also provide the specific processing choices made to run the DECam images that constitute DECaPS2. For each CCD in an exposure, we use the InstCal products from the DECam NOAO community pipeline (CP), which include the image, weight map, and data-quality mask. Since this is an afterburner code, we also take as inputs the \texttt{crowdsource} outputs for these images: the position and flux of sources, the background and background+sources model images, and the position-dependent PSF model. 

To create the initial mask for pixels that need infilling (see procedure in Section \ref{sec:prelim_infill}), we use the flux of each source and a static PSF model by evaluating the PSF at the center of the CCD. Then, pixels around the stellar position are masked if the \texttt{flux} $\times$ PSF $>$ \texttt{thr}, where \texttt{thr} is a chosen threshold. This threshold was chosen to mask residuals in the core of bright stars resulting from mismodeling. For \texttt{crowdsource} models of DECaPS2 images, we found \texttt{thr} $= 20$ ADU acceptable for all bands and fixed this parameter for all exposures. 

In addition, any pixel with a nonzero data-quality bit from the CP (cosmic rays, bleed trails, satellite trails, etc.) were masked. Finally, we masked pixels with residuals larger than \texttt{outthr} $= 20,000$ ADU that were previously unmasked. These generally resulted from distorted PSFs near the edge of CCDs that were not captured in the per-star masking (see Section \ref{sec:CompCost} for more). In the full DECaPS2 run, less than 1\% of the images had to use the infill fallback of the image median for some subset of the pixels. After infilling, we add Poisson noise to each pixel in the mask as in Equation \ref{eq:infill_draws}, where the gain used is that estimated from \texttt{crowdsource} as the median of the weight map $\times$ image.\footnote{Random steps use exposure date-time seeds for reproducibility.}

To handle modeled sources near or beyond the edge, the image is padded using reflective boundary conditions. Then the image is multiplied with a shifted version of itself, averaged with a boxcar mean ($N_x^{\rm{sample}}=N_y^{\rm{sample}}=129$), and stored as in Equation \ref{eq:boxcar}. For each source in the image, the covariance matrix for an $N_p = 33$ subimage is then constructed out of elements of those image products. The pixels in that subimage are then classified as ``good'' ($k$-pixels), ``hidden'' ($k_{\star}$-pixels), or ``ignored'' based on the PSF for a source at that location and the flux. All pixels in the PSF above \texttt{thr} (same value as during infill) are selected as ``hidden.'' If the star has flux $< 10,000$ ADU, it is masked as if the flux were $10,000$ ADU in order to ensure that the background behind every star is predicted.

Pixels masked by data-quality flags or outside a radius of 16 pixels that are not already ``hidden'' are ``ignored'', and all remaining pixels are selected as ``good.'' By making the choice of $N_p = 33$, we implicitly impose a cutoff on the brightest star we can model well. For stars so bright that the whole $N_p$ subimage is ``hidden,'' for example, there are no ``good'' pixels to use for the conditional prediction. However, the absolute uncertainty on flux calibration is usually the dominant term in the error budget for bright stars. So if the number of ``good'' pixels is too small, we compromise. The mask of all pixels beyond radius 16 is removed and the row of pixels around the outer edge of the subimage are forced to be ``good'' so there is at least a one-pixel deep boundary for conditioning. This fallback changes the value of our quality flag \texttt{dnt} from 0 to 2.\footnote{For heavily saturated stars, neither sky-subtraction nor deblending noticeably change the predicted flux.} A table of \texttt{dnt} bit values is available in Section \ref{sec:dataavil}.

\begin{figure}[t]
\centering
\caption{\textit{Top}: Residuals from \texttt{crowdsource} model of the central r-band region of Figure \ref{fig:CED116im}. \textit{Middle}: Z-scores for injected sources as a function of position for the image above. Clear spatial correlation between the sign and magnitude of the Z-scores and residuals is observed (correlation between \textit{top} and \textit{middle} panels). \textit{Bottom}: After LPI, the large population of $|$Z-scores$|$ larger than $3$ are removed (the scatter points are whiter) and little spatial correlation is observed.}
\label{fig:CED116}
\end{figure}

\begin{figure}
\centering
\includegraphics[width=\linewidth]{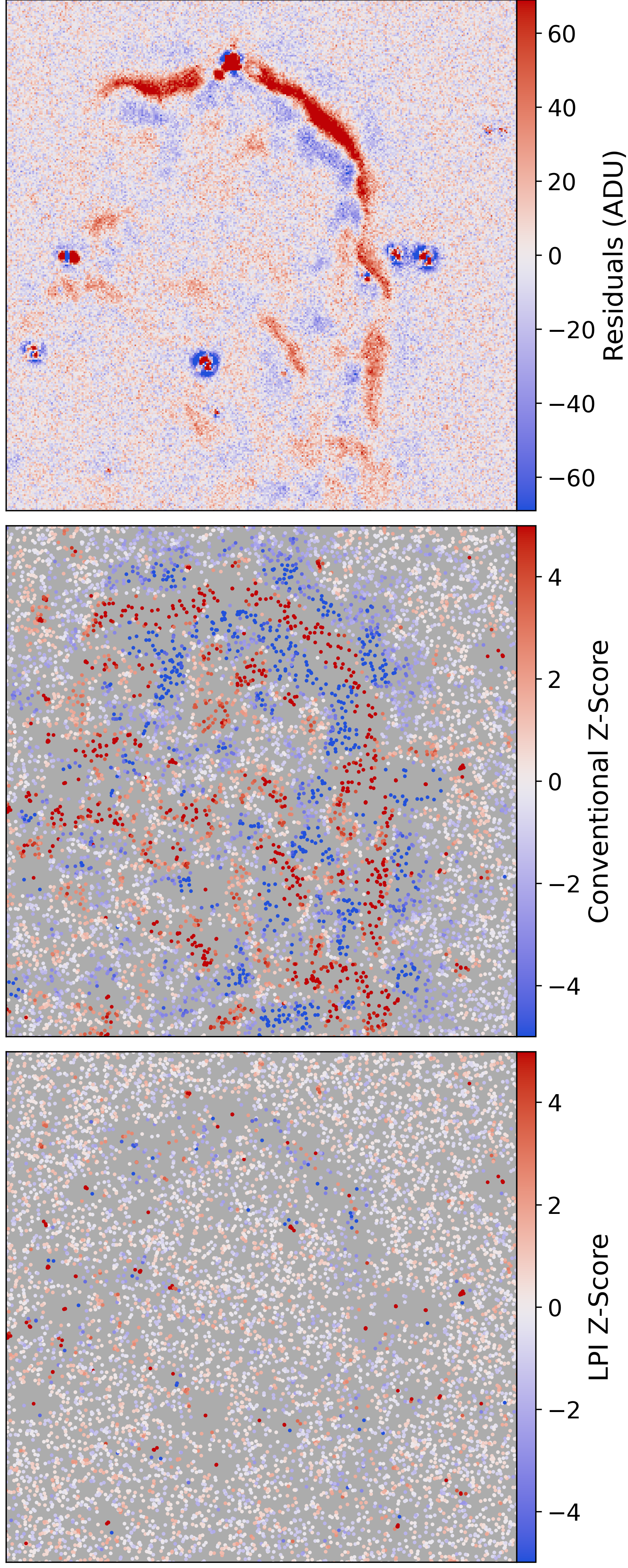}
\label{fig:CED116p}
\end{figure}

\begin{figure*}[ht]
\centering
\includegraphics[width=\linewidth]{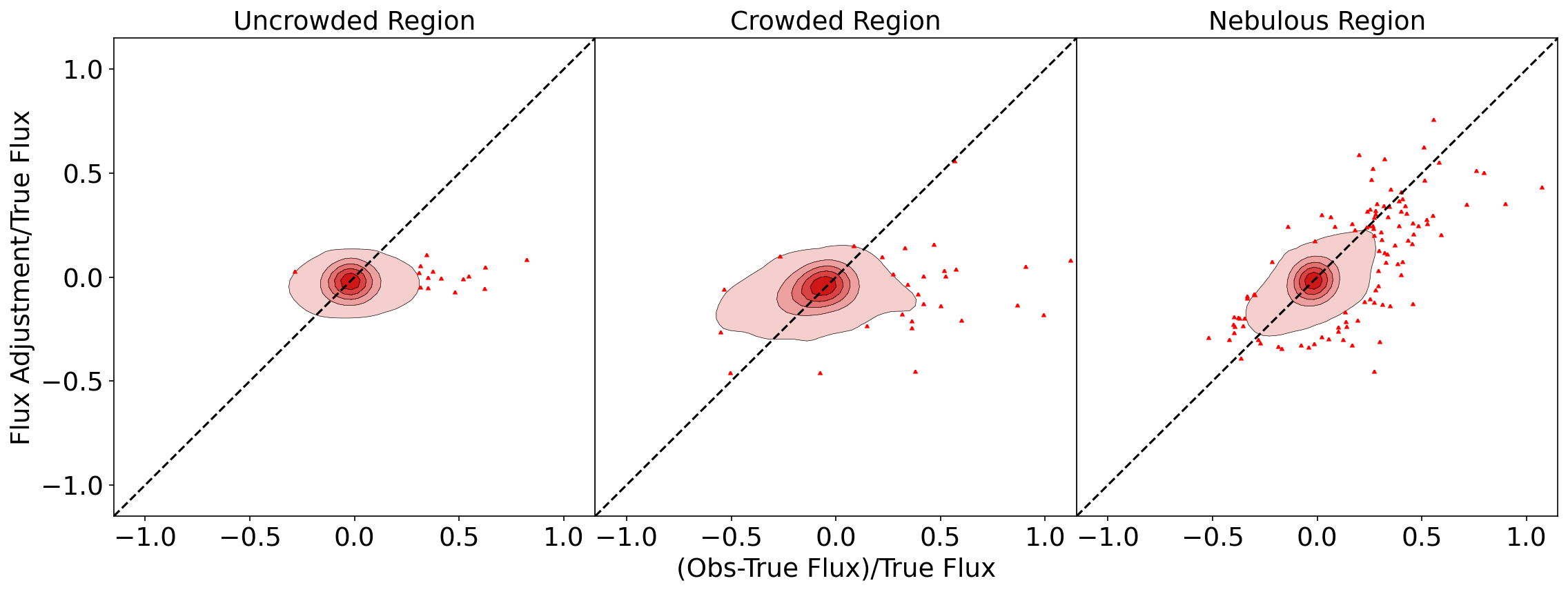}
\caption{The ratio of $\rm{f}^{\rm{b}}$ over true flux versus the fractional error in flux estimation for injected sources using \texttt{crowdsource}. The colored contours represent density for the 2D histogram, and sources beyond the lowest contour are shown as scatter points. Sources that fall along the 1:1 line (black dashed) are correctly debiased by LPI, points that fall along the $y = 0$ line are misestimated by \texttt{crowdsource} and not corrected, and points that fall along the $x = 0$ line are incorrectly modified by LPI. \textit{Left}: Almost no correlation is observed for typical uncrowded fields and no outliers are introduced by the debiasing. \textit{Middle}: For crowded regions, a slight correlation is observed and, while there is larger scatter, no outliers are introduced by debiasing. \textit{Right}: For the nebulous region, there is a strong positive correlation (contours follow the 1:1 line), showing that the LPI analysis improves the flux estimate.}
\label{fig:ScatterPError}
\end{figure*}

The PSF models from \texttt{crowdsource} can occasionally contain negative pixels, though usually only for non-photometric exposures.\footnote{Before any photometric cuts, these detections comprise less than 0.02\% of the total number of detections for DECaPS2.} To handle these cases gracefully, the min/max ratio of the PSF is computed and if the PSF is negative, bit 3 of \texttt{dnt} is set. To quantify the negative portion of the PSF, bit 4 and 5 are set if the ratio is $<-1 \times 10^{-3}$ and $-1 \times 10^{-1}$, respectively. In all cases, the absolute value of the PSF is used by \textsc{CloudCovErr.jl}.
\vspace{-2mm}
\begin{deluxetable}{ll}[hb]
\tablenum{1}
\tablecaption{\textsc{CloudCovErr.jl Outputs
\label{tab:outputs}}}
\tablecolumns{2}
\tablewidth{0pt}
\tablehead{
\colhead{Variable} & \colhead{Description}
}
\startdata
\texttt{dcflux} & flux uncertainty \\ 
\texttt{dcflux\_diag} & diagonal flux uncertainty \\
\texttt{cflux} & LPI corrected flux \\
\texttt{fdb\_res} & bias offset from residuals \\
\texttt{fdb\_pred} & bias offset from predicted background \\
\texttt{cchi2} & ``good'' pixel $\chi^2$ \\
\texttt{kcond0} & initial number of ``good'' pixels \\
\texttt{kcond} & number of ``good'' pixels \\
\texttt{kpred} & number of ``hidden'' pixels \\
\texttt{dnt} & quality flag on solution \\
\enddata
\end{deluxetable} \vspace{-13mm}

The last step before computing Equations \ref{eq:debiasFactors} and \ref{eq:debiasError} is adjusting the per-pixel variance (the diagonal of the local covariance matrix) to account for the Poisson noise from the presence of the bright star. The local covariance matrix was originally estimated only on pixels that were representative of uncontaminated background and thus only have Poisson noise on the order of the background sky counts. Thus we add the background+sources model minus the background model to the diagonal. Since covariance matrices must be symmetric positive semidefinite, we must never add a negative to the diagonal during this procedure. However, negative PSF models or sources modeled with negative flux can make this difference negative in rare cases, so we add the absolute value.

As a final quality-assurance check, we compute the $\chi^2$ for the ``good'' pixels given the local covariance matrix estimate. This provides a self-consistency check on the quality of the local covariance matrix estimate. Since the ``good'' pixels were used for conditioning the prediction on the ``hidden'' pixels, assuming the local covariance matrix was a decent model, large $\chi^2$ (relative to the number of degrees of freedom) act as a red flag.\footnote{In section \ref{sec:Toy}, we observed that the Gaussian model need not be a good representation of the overall population. Thus we cannot expect these values to follow a $\chi^2$ distribution exactly. However, values which are orders of magnitude too high likely indicate numerical instability.} The output catalogue columns are described in Table \ref{tab:outputs}. The number of ``ignored'' pixels is obtained as $N_p^2 -$ \texttt{kcond} $-$ \texttt{kpred}.

\subsection{Nebulous Region} \label{sec:Neb}

First we consider a r-band image of a highly structured \HII region (CED 116 near Lambda Centauri) in the survey footprint (Figure \ref{fig:CED116im}). We then create many (2000) new images by injecting synthetic sources using the PSF model derived during the \texttt{crowdsource} fit of the image. Each injection image consists of a number of synthetic sources which corresponds to 10\% of the number of sources originally found so as to not significantly perturb the image. These sources are also sampled from the distribution of fluxes in the original solution to the image, excluding sources with ``bad'' flags set or brighter than 17$^{\rm{th}}$ magnitude in g-band. The injected source-center positions are uniformly random, excluding $N_p = 33$ pixels around the edge of the CCD. The injection images are reprocessed by \texttt{crowdsource} as if they were standard images so we can compute Z-scores for the photometric solutions relative to the injected fluxes.

By plotting the residuals (Figure \ref{fig:CED116}, \textit{top}) and the original Z-scores (Figure \ref{fig:CED116}, \textit{middle}), we visualize a clear spatial correlation between the sign and magnitude of the Z-scores and the residuals of the \texttt{crowdsource} models. These correlations between structured backgrounds and biased photometry with underestimated error bars significantly complicate large-scale inference from photometry in the Galactic plane, such as for efforts mapping the interstellar dust distribution \citep{Green:2019:ApJ:}. After adding our flux bias offset and using the error bars that include off-diagonal contributions, we obtain the bottom panel of Figure \ref{fig:CED116} where there is no clear correlation of the Z-scores with the residuals and the excessive population of $>\pm3\sigma$ outliers has been removed. We take this as strong evidence in support of using our post-processing code with other/future photometric pipelines.

\subsection{Correction Correlations} \label{sec:CorectCorr}

We also want to confirm that using the LPI bias offset does not strongly affect regions without structured backgrounds. Since we have already shown that LPI improves over the conventional PDF (see Figure \ref{fig:crowdedflat}), we will examine the per-star percent changes. This allows us to confirm that LPI is not creating large flux errors in the process of debiasing. To do this we use injection tests as in Section \ref{sec:Neb}, with an additional requirement that the \texttt{dnt} quality flag is zero. 

The first panel of Figure \ref{fig:ScatterPError} shows the per-star percent change versus the percent error in a field of typical stellar density ($\sim14,000$ sources) without structured background.\footnote{The classification of a region as having a structured background is somewhat subjective. In the context of DECaPS, this is automated using a neural network to flag regions as ``nebulous'' or not. \citep{Schlafly:2018:ApJS:}} The maximum percent change is 20\% due to the bias offset. If the bias introduced by structured backgrounds were the main source of error, we would expect the per-star corrections to cluster along the 1:1 line (black dashed). However, the distribution appears flat and parallel to the x-axis. This agrees with the intuition that in uncrowded regions, errors from structured backgrounds are subdominant (to deblending errors for example). 

In the second panel, we compare to a crowded field ($\sim195,000$ sources). The scatter introduced by the bias offset is larger, up to 50\% in the most extreme cases, but the majority of stars are clustered near the origin. Only a slight positive trend is observed and might be attributed to the larger residuals in crowded fields. Finally, in a nebulous field (specifically g-band of the top panel of Figure \ref{fig:CED116}), we see a strong positive correlation between the correction factors and the conventional percent differences, as expected. Thus, we correct for bias resulting from structured background when present, but do not introduce bias into the catalogue on either flat or crowded regions.

Another metric to examine is  following ratio.

\begin{ceqn}
\begin{align} \label{eq:ratiostd}
\sigma\left(\frac{f^{\rm{b}}}{\texttt{dcflux}}\right)\bigg/\sigma\left(\frac{\hat{f}-f}{\texttt{dcflux}}\right)
\end{align}
\end{ceqn}
If the numerator is small, then the change in flux from the bias offsets is small relative to estimated uncertainty. We compare to the standard deviation of Z-scores to account for the fact that the distributions of uncorrected flux estimates for the three cases in Figure \ref{fig:ScatterPError} have different widths. For the uncrowded, crowded, and nebulous regions, this metric is $0.23$, $0.76$, and $0.95$, respectively. In the uncrowded field, this further demonstrates that the bias offsets do not significantly change the flux estimates relative to the flux uncertainties. And, in the nebulous field, the bias offsets are large, as expected.

\begin{figure}
\centering
\includegraphics[width=\linewidth]{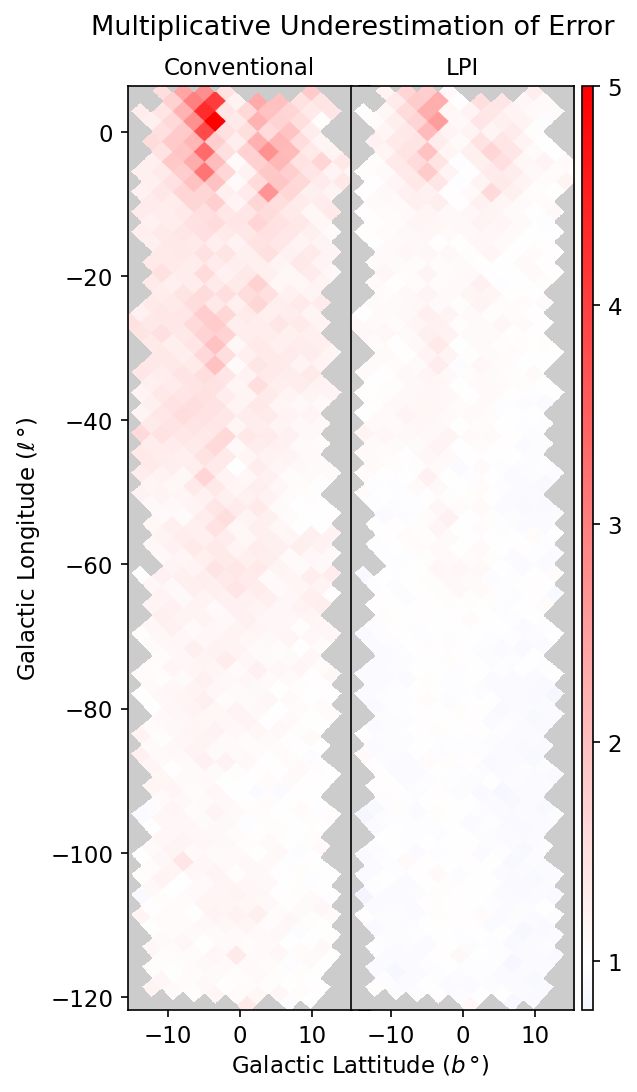}
\caption{Factor by which the estimated flux uncertainty is too small for conventional (\textit{left}) and LPI photometry (\textit{right}). In the ideal case, this factor would be 1 (white) everywhere. The multiplicative factor is computed as IQR/1.34896 for the Z-scores of injected sources, where the denominator is the IQR for the standard normal distribution. The map is over the survey footprint of DECaPS2 at \texttt{NSIDE} = 32. LPI photometry is a clear improvement, though correction toward the Galactic bulge is imperfect.}
\label{fig:IQR}
\end{figure}

\subsection{Full Survey Footprint} \label{sec:AllSky}

Zooming out, we evaluate the performance of the LPI error bars on the entire survey footprint (2700 deg$^2$, 6.5\% of the sky). In the ideal case, we expect the Z-scores to be normally distributed when the uncertainties are correctly estimated. However, biases and failure modes in photometric pipelines introduce outliers. We measure the interquartile range (IQR), which is robust to outliers, for both the conventional and LPI Z-scores and compare this to the ideal value for a Gaussian distribution of 1.349. Then, we report by what scalar the error bars must be multiplied to achieve the IQR of a normal distribution, shown in Figure \ref{fig:IQR} (\texttt{NSIDE}=32). To do this, we use the injection tests (with the same criterion as before) and only analyze Healpix pixels with $>30$ stars.

In the conventional case, the rescaling required is over a factor of 5 toward the Galactic bulge. In contrast, the rescaling for the LPI error bars is only a factor of 2 toward the Galactic bulge. Comparing the median rescaling factors, we have 1.17 and 1.02 for the conventional and LPI error bars, respectively. To have a correct central value for the rescaling ($1.0$), there must be (and are) cases where the rescaling is less than one. So the ideal case for Figure \ref{fig:IQR} is on average white with small red and blue fluctuations, as is seen for the majority of the LPI panel. Deviations from ideality toward the Galactic bulge likely result from the large masking fraction in this region of high stellar density, which impedes estimation of the residual covariance.

While it may be tempting to simply rescale the conventional estimates and then avoid the computational cost associated with LPI, a simple rescaling of error bars will never be able to remove the spatial correlations shown in Figure \ref{fig:CED116}. Just as in the conventional case, these rescalings indicate a gap between the model complexity and reality that requires further method development to close. In the meantime, we leave it to users to decide whether to rescale the LPI uncertainty estimates.

We can evaluate relative metrics, such as the ratio between the stated flux uncertainty from LPI and \texttt{crowdsource}, with higher angular resolution (Figure \ref{fig:ErrRatio}, \texttt{NSIDE}=512).\footnote{We report an outlier-clipped mean value of the ratio for stars within each Healpix pixel. The outliers are beyond 10*IQR/1.349 and are computed per ccd. In addition to requiring \texttt{dnt} = 0, we impose \texttt{cchi2}/\texttt{kcond} $> 0.7$ from the \textsc{CloudCovErr.jl} flags. We also require no ``bad flags'' (which removes detections on CCD S7), \texttt{rchi2} $< 100$, and \texttt{fracflux} $> 0.5$ on the \texttt{crowdsource} outputs.} The main feature is a large (relative) increase in the LPI error bars toward the Galactic bulge. This correlation with stellar density is consistent with intuition developed for the crowded-field test-case (Figure \ref{fig:crowdedflat}). In addition, the peak associated with the Carina Nebula ($b = -0.8\degree, \ell = -73\degree$) and filaments associated with the Vela supernova remnant ($b = -2.8\degree, \ell = -96\degree$) shows that the LPI errors are sensitive to structured backgrounds as intended. More detailed analysis of these and other metrics over the survey footprint will be part of the DECaPS2 release \citep{Saydjari:2022:ip:}.

\begin{figure} [t!]
  \caption{Ratio of uncertainty estimates for LPI relative to conventional photometry over the full survey footprint of DECaPS2. A value of 1 indicates LPI and conventional photometry agree on the uncertainty estimate. Features associated with the Galactic bulge, Carina Nebula ($b = -0.8\degree, \ell = -73\degree$), and Vela supernova remnant ($b = -2.8\degree, \ell = -96\degree$)  stand out.}
  \label{fig:ErrRatio}
\end{figure}

\section{Computational Details} \label{sec:CompCost}

\begin{figure}
\centering
\includegraphics[width=0.98\linewidth]{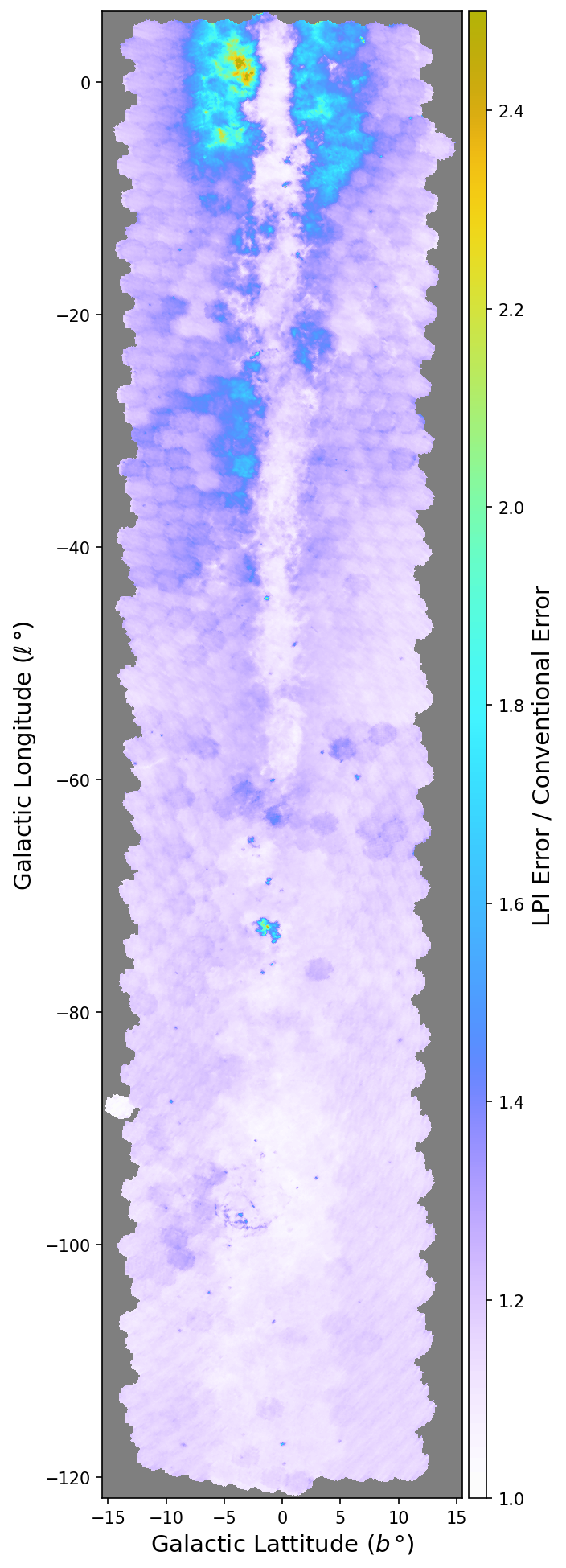}
\label{fig:ErrRatiop}
\end{figure}

In our LPI photometry implementation (see Section \ref{sec:dataavil}), for a field of average stellar density in DECaPS ($\sim10,000$ sources/8 Megapixels) the majority of the computational time is split evenly between elementwise image multiplication (Hadamard products) and per-star Cholesky decomposition. We choose to construct the local covariance matrix from elements of the boxcar mean of the initial image multiplied by a shifted version of that image (Equation \ref{eq:boxcar}). For 2D, this costs $\sim 4N_p^2N_xN_y$ FLOPs in contrast to constructing $V^TV$ via matrix multiplication per star which is $\sim 2N_{\rm{star}}N_p^4N_x^{\rm{sample}}N_y^{\rm{sample}}$. Here $N_x$ and $N_y$ are the size of the original image and $N_x^{\rm{sample}}$ and $N_y^{\rm{sample}}$ are the size of the boxcar window in each direction.

There can be significant memory requirements to holding those image products in storage for use in the per-star covariance matrix construction. The alternative consists of grabbing the needed elements of the product images for every star at each shift, and throwing away the product image. However, this alternative procedure requires highly variable amounts of memory that is strongly dependent on the number of stars in the exposure. We believe in the context of large survey pipelines, stable memory requirements are highly desirable as they maximize the CPU utilization.

In order to not store $\sim 2 N_p^2 N_x N_y $ \texttt{Float32}s during the processing of each CCD, $\sim 70$ GB for DECam with $N_p = 33$ as in Section \ref{sec:Survey}, we break each image into tiles and process each tile and the stars within it separately. Doing so incurs an overhead cost since we have to include some padding beyond the furthest stellar position in a given tile. This tiling procedure allowed us to run the full DECaPS2 survey with $3.6-4.8$ GB per exposure (median 3.7 GB) in 203k core-hours. The jobs were run on the FASRC Cannon cluster at Harvard University on an entire compute node with 2 water-cooled Intel 24-core Platinum 8268 Cascade Lake CPUs and 192GB RAM running 64-bit CentOS 7. The runtime versus number of stars for the DECaPS2 run demonstrates high linearity with respect to the number of stars per exposure-- approximately 15.3 core-ms/star in addition to a constant per exposure ($\sim 62$ CCDs) of 3.4 core-h (see \cite{Saydjari:2022:ip:} for more details).

The major failure mode for the code presented in this work occurs when numerical effects cause the estimated covariance matrix to not be positive definite, which must be true theoretically for any covariance matrix (and prevents Cholesky factorization). The main sources of instability we find are unmasked bright pixels which are not well captured by the \texttt{crowdsource} model. These outliers prevent a positive definite estimate of the covariance matrix and should be masked anyway as they do not represent the true background. 

In DECaPS2, we reduced \texttt{outthr} to $5000$ and reran the 980 CCDs (out of 1.32 million CCDs) which contained stars without positive definite covariance estimates. Most of these resulted from bright stars not fully masked by our procedure because \textsc{crowdsource} modeled them as multiple fainter sources. Afterwards, only 9 CCDs in the whole survey contain stars with non-positive definite covariance estimates, and the majority of those result from large pupil ghost artifacts. In these cases where the source and sky model coming from \texttt{crowdsource} and the mask from the DECam NOAO community pipeline are all wrong, we simply leave all \textsc{CloudCovErr.jl} outputs as \texttt{NaN} as a warning to users. Using \texttt{Float64} instead of \texttt{Float32} (as we have done in all presented results) does enhance the numerical stability, but at too large a computational and memory cost.

Looking toward other large survey applications targeting the galactic plane, such as Rubin and Roman, we comment that the code provided in Section \ref{sec:dataavil} is not fully optimized. In several places, memory is re-allocated (which is inefficient) or the minimum memory footprint could be reduced by restructuring the code. Thus, the benchmarks above are not meant to demonstrate the optimality of the code, merely that estimating a covariance matrix and Cholesky-factorizing it for 34 billion detections can be done faster than the original photometric solutions in crowded fields. For instance, the \texttt{crowdsource} solutions for DECaPS2 required 230k core-hours (13.1 kpix/s). Of course, a GPU-based implementation could see significant speed-ups on the limiting steps of elementwise image multiplication and Cholesky factorization \citep{naumov2021incomplete}, but large surveys will have to weigh the relative cost of GPU/CPU resources to the relative speed-ups.

\section{Conclusion} \label{sec:Conc}

In this work, we have derived a correction to both the estimated flux and flux uncertainties by using local pixelwise infilling (LPI) to estimate the background behind a star. We showed that these corrections improve photometry on synthetic and real fields with structured backgrounds. In particular, we applied our method to all 34 billion detections in the DECaPS2 survey, demonstrating scalability and improvement across the entire survey footprint. In contrast to the commonly used method of rescaling error bars to obtain the expected chi-square, we showed that LPI adjusts uncertainty estimates according to the local background covariance, and correctly treats both positive and negative residuals. In the crowded limit, we found both improved estimation of contamination of the sky background by unmodeled stellar flux and uncertainty estimates from large residuals. We showed that LPI reduces to a conventional uncertainty (diagonal covariance) on fields without structured backgrounds, allowing for uniform processing of an entire photometric catalogue. In addition, we point out that LPI applies to a broader class of images beyond Gaussian fields, even capturing highly non-Gaussian structures such as isolated filaments and clouds. We demonstrate this both analytically and numerically on a real dust image. Finally, we release our code base which is a pure afterburner to any photometric code that outputs residual images and a PSF model. Thus, we suggest that this method can be easily used to create value-added photometric catalogues for existing and future surveys in the Galactic plane.

While we restricted this work to point sources, the same technique should be extensible to correcting estimates of the flux from more extended sources, such as galaxies. In that case, the shape function for the extended source (i.e. the PSF-convolved Sérsic profile for a galaxy) takes the role of the point spread function in the above discussion. Future directions include demonstrating that these corrections improve downstream inference, e.g Bayesian fitting of photometry to jointly predict intrinsic stellar parameters (temperature, age, etc.) and foreground dust densities as a result of reddening. We expect that the improved flux uncertainties should expand the set of well-modeled sources. This is especially important for dust mapping in the Galactic plane, where there is significant extinction, and the stars in fields with structured backgrounds are precisely those most constraining for the dust clouds and filaments of interest.
\vspace{-3mm}

\section{Code and Data Availability} \label{sec:dataavil}

Data products and \textsc{jupyter} notebooks required to reproduce figures within the paper are publicly available on \href{https://doi.org/10.5281/zenodo.5809522}{\textbf{Zenodo}} (11.0 GB). We make the code base used here publicly available on GitHub \href{https://github.com/andrew-saydjari/CloudCovErr.jl}{\textbf{CloudCovErr.jl}} and used v0.9.0 for this work. The data products associated with the full DECaPS2 run will be described in more detail and accompany the survey release paper.
\vspace{-6mm}

\acknowledgments
A.S. gratefully acknowledges support by a National Science Foundation Graduate Research Fellowship (DGE-1745303). D.F. acknowledges support by NSF grant AST-1614941, “Exploring the Galaxy: 3-Dimensional Structure and Stellar Streams.” This work was supported by the National Science Foundation under Cooperative Agreement PHY-2019786 (The NSF AI Institute for Artificial Intelligence and Fundamental Interactions). We would like to thank Edward Schlafly and Lucas Janson for feedback during the development of this project and writing of this manuscript. We acknowledge Eugene Magnier, Josh Speagle, Catherine Zucker, Justina R. Yang, Nayantara Mudur, and Christopher Stubbs for helpful discussions. A.S. acknowledges Sophia S\'{a}nchez-Maes for helpful discussions and much support. Computations in this paper were run on the FASRC Cannon cluster supported by the FAS Division of Science Research Computing Group at Harvard University. We acknowledge the developers and maintainers of the Julia ecosystem, from whom we have learned and benefited greatly.

This project used data obtained with the Dark Energy Camera (DECam), which was constructed by the Dark Energy Survey (DES) collaboration. Funding for the DES Projects has been provided by the U.S. Department of Energy, the U.S. National Science Foundation, the Ministry of Science and Education of Spain, the Science and Technology Facilities Council of the United Kingdom, the Higher Education Funding Council for England, the National Center for Supercomputing Applications at the University of Illinois at Urbana-Champaign, the Kavli Institute of Cosmological Physics at the University of Chicago, the Center for Cosmology and Astro-Particle Physics at the Ohio State University, the Mitchell Institute for Fundamental Physics and Astronomy at Texas A\&M University, Financiadora de Estudos e Projetos, Fundação Carlos Chagas Filho de Amparo à Pesquisa do Estado do Rio de Janeiro, Conselho Nacional de Desenvolvimento Científico e Tecnológico and the Ministério da Ciência, Tecnologia e Inovacão, the Deutsche Forschungsgemeinschaft, and the Collaborating Institutions in the Dark Energy Survey. The Collaborating Institutions are Argonne National Laboratory, the University of California at Santa Cruz, the University of Cambridge, Centro de Investigaciones Enérgeticas, Medioambientales y Tecnológicas-Madrid, the University of Chicago, University College London, the DES-Brazil Consortium, the University of Edinburgh, the Eidgenössische Technische Hochschule (ETH) Zürich, Fermi National Accelerator Laboratory, the University of Illinois at Urbana-Champaign, the Institut de Ciències de l’Espai (IEEC/CSIC), the Institut de Física d’Altes Energies, Lawrence Berkeley National Laboratory, the Ludwig-Maximilians Universität München and the associated Excellence Cluster Universe, the University of Michigan, the NOIRLab (formerly known as the National Optical Astronomy Observatory), the University of Nottingham, the Ohio State University, the University of Pennsylvania, the University of Portsmouth, SLAC National Accelerator Laboratory, Stanford University, the University of Sussex, and Texas A\&M University.

\software{
GNU Parallel \citep{Tange:2020:zndo:},
\textsc{astropy} \citep{AstropyCollaboration:2013:A&A:},
\textsc{h5py} \citep{collette_python_hdf5_2014}, 
\textsc{ipython} \citep{Perez:2007:CSE:}, 
\textsc{matplotlib} \citep{Hunter:2007:CSE:}, 
\textsc{numpy} \citep{vanderWalt:2011:CSE:}, 
\textsc{scipy} \citep{Virtanen:2020:NatMe:}
}
\appendix
\vspace{-5mm}
\section{Pixel-Space Linear Least Squares Notation}\label{sec:LS}

To reformulate our discussion in the language of a linear least squares regression in pixel-space, we view the ``good'' pixels ($k$) as the regressors for the value of the ``hidden'' pixels ($k_\star$), the dependent variables. Let $X$ and $Y$ be matrices of the regressors and dependent variables, respectively, where each row represents a different sample (obtained by pixel shifts over the local region) out of a total of $N$ samples. In what follows, we will ignore centering of the data for clarity. Then, the predicted linear parameters are given as usual.

\begin{ceqn}
\begin{align} \label{eq:beta}
\hat \beta = (X^TX)^{-1}X^TY
\end{align}
\end{ceqn}

The prediction of the ``hidden'' pixels ($\hat y$) given the observed ``good'' pixels for a given stamp ($x_0$) is then

\begin{ceqn}
\begin{align} \label{eq:pred}
\hat{y} = \hat{\beta}^T x_0 = Y^TX\left(X^TX\right)^{-1}x_0
\end{align}
\end{ceqn}

The covariance of this prediction is just

\begin{ceqn}
\begin{align} \label{eq:preduncert}
\mathbb{E}\left[(y-\hat{\beta}^Tx)(y-\hat{\beta}^Tx)^T\right] = \frac{1}{N-1}\left(Y^TY-Y^TX\left(X^TX\right)^{-1}X^TY\right)
\end{align}
\end{ceqn}

where the expectation value is estimated over the $N$ samples. Then the correspondence to the Gaussian process regression notation used in the main text is clear after identifying $x_0 = X_k$, $\hat{y} = \bar{X}(k_\star|k)$, and partitioning $V$ into $k$ and $k_\star$ sub-blocks so that we can see

\begin{ceqn}
\begin{align} \label{eq:LSeq}
C = \frac{1}{N-1} V^TV = \frac{1}{N-1}
\begin{bmatrix}
X^T\\
Y^T
\end{bmatrix}
\begin{bmatrix}
X & Y\\
\end{bmatrix}
= \frac{1}{N-1}
\begin{bmatrix}
X^TX & X^TY\\
Y^TX & Y^TY
\end{bmatrix}
=
\begin{bmatrix}
C_{k,k} & C_{k,k_\star}\\
C_{k_\star,k} & C_{k_\star,k_\star}
\end{bmatrix}
\end{align}
\end{ceqn}

\begin{figure}[b]
\centering
\includegraphics[width=\linewidth]{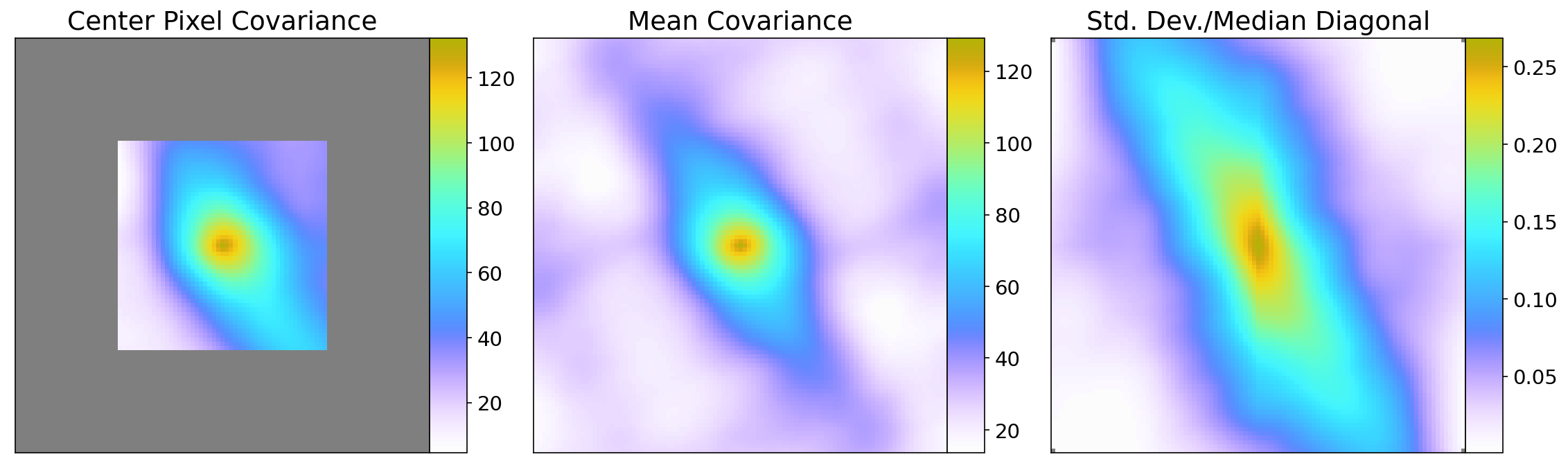}
\caption{Covariance function for reconstruction of a portion of the WISE 12 $\mu$m dust map. \textit{Left}: Covariance of the central pixel with all other pixels in the $49 \times 49$ pixel subimage. \textit{Middle}: Mean value of the covariance as a function of pixel displacement for all pairs of pixels in the subimage. \textit{Right}: Standard deviation of the covariance between pixel pairs with the same displacement normalized by the median value of the diagonal (pixelwise variance). This demonstrates that the LPI covariance function is stationary at the $\sim 10\%$ level outside the core, and $\sim 25\%$ level in the core.}
\label{fig:CovStation}
\end{figure}

\section{Approximate Stationarity of LPI Covariance}\label{sec:station}

To show the approximate stationarity of the covariance matrix estimated in LPI, we consider the reconstruction of a portion of the WISE 12 $\mu$m dust map (Section \ref{sec:WISE}). This test represents an extreme in heterogeneity of the local background used for estimating the local covariance matrix. In Figure \ref{fig:CovStation} (\textit{left}) we show the covariance of the central pixel with all other pixels in the $49 \times 49$ pixel subimage. In Figure \ref{fig:CovStation} (\textit{middle}), we show the mean value of the covariance as a function of pixel displacement for all pairs of pixels in the subimage. Note that the central pixel covariance strongly resembles the mean covariance as a function of displacement, though there are some small distortions. More quantitatively, in Figure \ref{fig:CovStation} (\textit{right}) we show the standard deviation of the covariance between pixel pairs with the same displacement normalized by the median value of the diagonal (pixelwise variance). This demonstrates that the covariance function is stationary at the $\sim 10\%$ level outside the core, and $\sim 25\%$ level in the core. 

The mean covariance as a function of displacement is highly anisotropic, agreeing with our intuition that anisotropy is required to handle filamentary structures. A more rigorous justification of stationarity could use a covariance matrix based only on the mean covariance as a function of displacement to perform the reconstructions as in Figure \ref{fig:WISE} and determine if there is any measurable difference in the reconstructions compared to our LPI procedure. Further, except for low amplitude structure near the edges of Figure \ref{fig:CovStation} (\textit{middle}), the covariance function appears to decrease rapidly as a function of radius. This suggests that the covariance function might be well-approximated by a function such that the covariance matrix is positive definite when sampled at an arbitrary points (Bochner's theorem). Whether or not exact stationarity can be enforced or a well-behaved kernel function approximation found while maintaining sufficient flexibility to describe the diversity of structure in the ISM remains an exciting question for future work.

\begin{figure}[b]
\centering
\includegraphics[width=\linewidth]{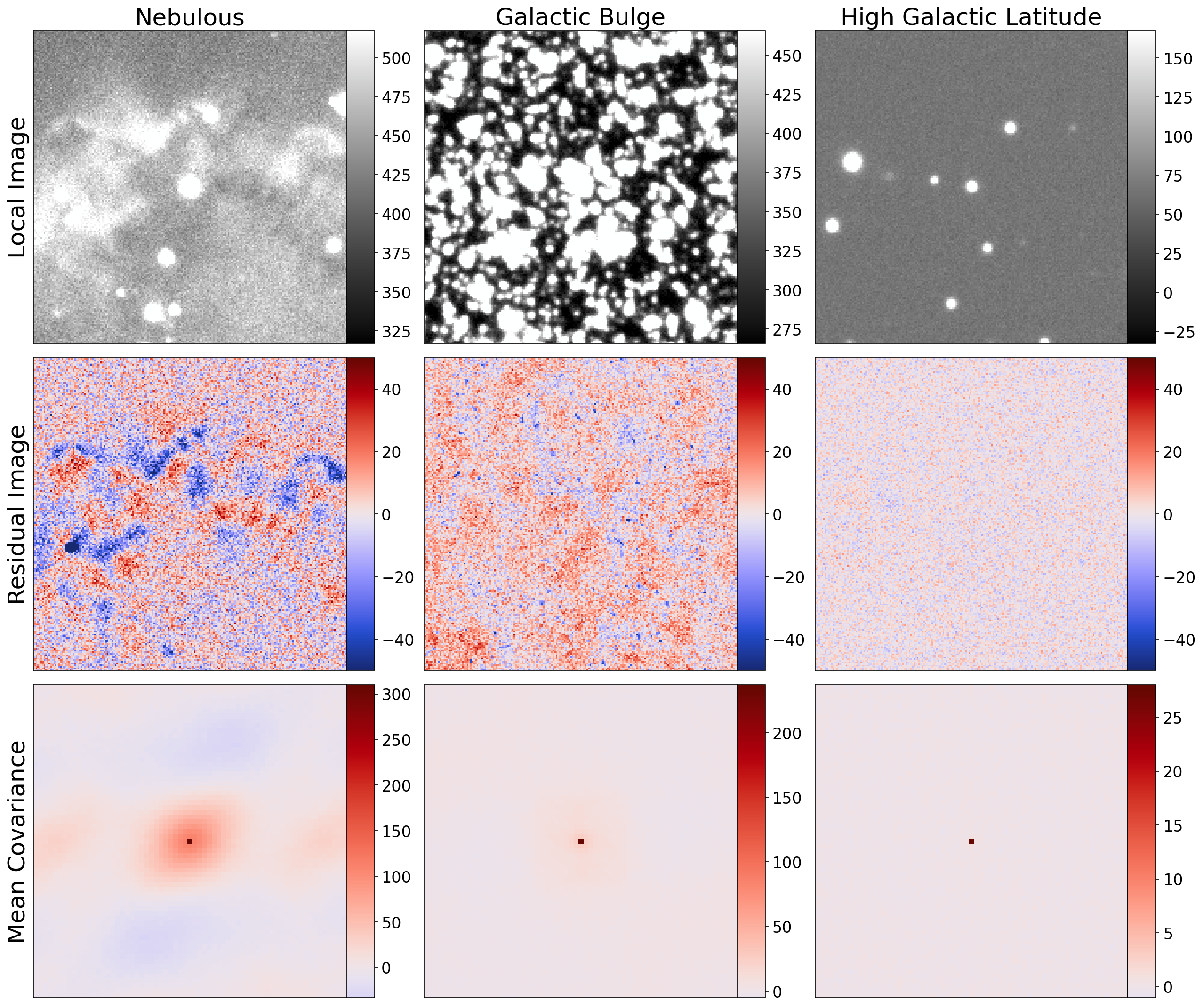}
\caption{Diversity in covariance structure obtained in different regions of the DECaPS2 survey footprint (all r-band images): a nebulous region (left column), a dense field in the Galactic bulge (center column), and a low density field at high Galactic latitude (right column). In the top row, the raw image of the star and surrounding field are shown. Each image is stretched $\pm50$ around its median. In the center row, we display the \texttt{crowdsource} residuals after infilling masked pixels as described in Section \ref{sec:prelim_infill}. The image spatial scale is chosen to include all training data used to estimate the local covariance matrix of the unmodeled background structure. In the bottom row, we display the mean stationary approximation (as in Figure \ref{fig:CovStation}, \textit{middle}) to the covariance of the residuals used in our LPI procedure.}
\label{fig:CovDiv}
\end{figure}

In Figure \ref{fig:CovDiv} we show a small sample of the diversity in covariance structure obtained in different regions of the DECaPS2 survey footprint (all r-band images): a nebulous region (left column), a dense field in the Galactic bulge (center column), and a low density field at high Galactic latitude (right column). In the top row, the raw image of the star and surrounding field are shown. Each image is stretched $\pm50$ around its median. In the center row, we display the \texttt{crowdsource} residuals after infilling masked pixels as described in Section \ref{sec:prelim_infill}. The image spatial scale is chosen to include all training data used to estimate the local covariance matrix of the unmodeled background structure. In the bottom row, we display the mean stationary approximation (as in Figure \ref{fig:CovStation}, \textit{middle}) to the covariance of the residuals used in our LPI procedure. 

For the nebulous region, the long-range correlations are comparable to the pixelwise noise (magnitude of the central pixel) and are notably anisotropic. In the Galactic bulge, the pixelwise noise is more dominant than long-range correlations. However, as evidenced by Figure \ref{fig:ErrRatio}, our covariance estimate in the bulge is imperfect. This is in part because our masking fraction in regions of high density is so large. Uncertainty due to residuals from mismodeling neighboring stars (which are currently masked to access the covariance of the background) is likely important in this limit and incorporating this uncertainty into the covariance estimate is a route for future work. At high Galactic latitudes, the uncertainty is dominated by the pixelwise uncertainty, as expected and as shown in Figure \ref{fig:FilTest}. While we have focused on r-band, the covariance functions for other bands are similar. The covariance function for the same region, but in different bands may differ as a result of varying SNR, seeing, artifacts from diffraction spikes, or the difference in backgrounds visible at those wavelengths.

\bibliography{CloudCovErr.bib}{}
\bibliographystyle{aasjournal}
\end{document}